\documentclass[a4paper,12pt]{article}
\usepackage{epsf,amsmath}
\usepackage[dvips,usenames]{color}
\usepackage{graphicx}
\usepackage{color}
\usepackage{colortbl}

\newlength{\dinwidth}
\newlength{\dinmargin}
\newcommand{\spur}[1]{\not\! #1 \,}

\newcommand{\half}{\frac{1}{2}}
\begin{document}
\title{\bf Revisiting $B\to\pi K, \pi K^{\ast}$ and $\rho K$ Decays: CP Violations and Implication for New Physics }
\author{Qin Chang$^{a,b}$, Xin-Qiang Li$^{c}\footnote{Alexander-von-Humboldt Fellow}$,
Ya-Dong Yang$^a $\footnote{Corresponding author}\\
{$^{a}$\small Institute of Particle Physics, Huazhong Normal University, Wuhan,
Hubei  430079, P. R. China}\\
{ $^b$\small Department of Physics, Henan Normal University,
Xinxiang, Henan 453007, P.~R. China}\\
{$^c$\small Institut f\"ur Theoretische Physik E, RWTH Aachen,
D-52056, Aachen, Germany}\\
%{\small E-mail: changqin@iopp.ccnu.edu.cn,
%xinqiang@physik.rwth-aachen.de, yangyd@iopp.ccnu.edu.cn}
}
\date{}
\maketitle
\bigskip\bigskip
\maketitle \vspace{-1.5cm}

\begin{abstract}
Combining the up-to-date experimental information on $B\to\pi K, \pi
K^{\ast}$ and $\rho K$ decays, we revisit the decay rates and CP
asymmetries of these decays within the framework of QCD
factorization. Using an infrared finite gluon propagator of Cornwall
prescription, we find  that the time-like  annihilation amplitude
could contribute a large strong phase, while the space-like hard
spectator scattering amplitude is real. Numerically, we find that
all the  branching ratios and most of the direct CP violations,
except $A_{CP}(B^{\pm}\to K^{\pm}\pi^{0})$,  agree with
the current experimental data with an effective gluon mass
$m_g\simeq0.5~{\rm GeV}$. Taking the  unmatched difference in direct
CP violations between $B\to\pi^{0} K^{\pm}$ and $\pi^{\mp}K^{\pm}$
decays as a hint of new physics, we perform a model-independent
analysis of new physics contributions with a set of
$\bar{s}(1+\gamma_{5})b\otimes\bar{q}(1+\gamma_{5})q$ (q=u,d)
operators. Detail  analyses of the relative impacts of the operators
are presented in  five cases. Fitting the  twelve  decay modes,
parameter spaces are found generally with nontrivial weak phases.
Our results may indicate that both strong phase from annihilation
amplitude and new weak phase from new physics are needed to resolve
the $\pi K$ puzzle.  To further test the new physics hypothesis, the mixing-induced 
CP violations in $B\to\pi^{0}K_{S}$ and $\rho^{0}K_{S}$ are discussed and good agreements 
with the recent experimental data are found. 

\end{abstract}
%\noindent {\bf PACS Numbers: 13.25.Hw, 12.38.Bx,12.15mm, 11.30.Hv.}

\newpage

\section{Introduction}
With the fruitful running of BABAR and Belle in past decade, plenty
of exciting results has been produced, which provides a very fertile
testing ground for the  Standard Model~(SM) picture of flavor
physics and CP violations. Although most of the measurements are in
perfect agreement with the SM  predictions, there still exist some
unexplained mismatches. Especially,  a combination of experimental
data on  a set of  related decays will increase the tension between
the SM predictions and experimental measurements.  At present, there
are discrepancies  between the measurement of several observables in
$B\to \pi K$ decays and the predications of the SM,  the so-called
``$\pi K$ puzzle"~\cite{pikpuz},
 which have attracted  extensive investigations in the
SM~\cite{Beneke2,Beneke3,YD,PikPQCD,PikSCET,Pik}, as well as with
various specific New Physics~(NP) scenarios~\cite{PikNP}.

 Recently, Belle has measured the direct CP violations $B\to K\pi$ decays~\cite{Belle_Nature}
\begin{eqnarray}\label{Acpbelle}
 A_{CP}(B^{-}\to K^{-}\pi^{0})&\equiv&
 \frac{\Gamma(B^{-}\to K^{-}\pi^{0})- \Gamma(B^{+}\to K^{+}\pi^{0})}
 {\Gamma(B^{-}\to K^{-}\pi^{0})+\Gamma(B^{+}\to K^{+}\pi^{0}) }=+0.07\pm0.03\pm0.01,  \\
 A_{CP}(\bar{B}^{0}\to K^{-}\pi^{+})&\equiv&
 \frac{\Gamma(\bar{B}^{0}\to K^{-}\pi^{+})- \Gamma(B^{0}\to K^{+}\pi^{-})}{
 \Gamma(\bar{B}^{0}\to K^{-}\pi^{+})+ \Gamma(B^{0}\to K^{+}\pi^{-}) }
 =-0.094\pm0.018\pm0.008.
 \end{eqnarray}
The difference between direct CP violations in charged and neutral modes is
\begin{eqnarray}\label{cpdiff}
\Delta A \equiv A_{CP}(B^{-}\to K^{-}\pi^{0})-  A_{CP}(\bar{B}^{0}\to K^{-}\pi^{+})=0.164\pm0.037.
\end{eqnarray}
The  averages of the current experimental data of BABAR~\cite{bar},
Belle~\cite{Belle_Nature}, CLEO~\cite{cleocp} and CDF~\cite{cdfcp}
by the Heavy Flavor Averaging  Group (HFAG)~\cite{HFAG} are
%%%%%%%%%%%%%%%%%%%%%%%%%%%%%%%%%%%%%%%%%%%%%%%%%
\begin{eqnarray}\label{AcpPuzzleexp}
 &&A_{CP}(B^{-}\to K^{-}\pi^{0})=0.050\pm 0.025~,\nonumber\\
 &&A_{CP}(\bar{B}^{0}\to K^{-}\pi^{+})=-0.097\pm 0.012,
 \end{eqnarray}
%%%%%%%%%%%%%%%%%%%%%%%%%%%%%%%%%%%%%%%%%%%%%%%%%
and the difference $ \Delta A=0.147\pm0.028$ is established at
$5\sigma$ level. However, within the SM, it is generally expected
that $A_{CP}(\bar{B}^0_d\to\pi^+ K^-) $ and $A_{CP}(B^-_u\to\pi^0
K^-)$ are close to each other. For example, the recent theoretical
predictions for these two quantities based on the QCD factorization
approach~(QCDF)\cite{Beneke1}, the perturbative QCD
approach~(pQCD)\cite{KLS} and the soft-collinear effective
theory~(SCET)~\cite{scet} read
%%%%%%%%%%%%%%%%%%%%%%%%%%%%%%%%%%%%%%%%%%%%%%%%%
\begin{eqnarray}
\label{AcpPuzzleQCD}
&&\left\{\begin{array}{l}
A_{CP}(B^-_u\to\pi^0 K^-)_{QCDF}=-3.6\%~, \\
A_{CP}(\bar{B}^0_d\to\pi^+ K^-)_{QCDF}=-4.1\%~;
\end{array}\right.~QCDF~ Scenario ~S4~\cite{Beneke3}\\
\label{AcpPuzzlePQCD} &&\left\{\begin{array}{l}
A_{CP}(B^-_u\to\pi^0 K^-)_{PQCD}=(-1^{+3}_{-5})\%~, \\
A_{CP}(\bar{B}^0_d\to\pi^+ K^-)_{PQCD}=(-9^{+6}_{-8})\%~;
\end{array}\right.~pQCD~\cite{PikPQCD}\\
\label{AcpPuzzleSCET} &&\left\{\begin{array}{l}
A_{CP}(B^-_u\to\pi^0 K^-)_{SCET}=(-11\pm9\pm11\pm2)\%~, \\
A_{CP}(\bar{B}^0_d\to\pi^+ K^-)_{SCET}=(-6\pm5\pm6\pm2)\%.
\end{array}\right. SCET~\cite{PikSCET}
\end{eqnarray}
%%%%%%%%%%%%%%%%%%%%%%%%%%%%%%%%%%%%%%%%%%%%%%%%%
We can see that the present theoretical estimations within the SM
are  confronted with the established $\Delta A$. The mismatch may be
due to our limited understanding of the strong dynamics in B decays
which hinders precise estimations of the SM contributions, but
equally possible due to new physics effects~\cite{peskin, mannel}.

As is known, the annihilation decay of B meson into two light mesons
offers interesting probes for the dynamical mechanism governing
these decays, as well as the exploration of CP violation. In most of
B meson non-leptonic decays, the annihilation corrections could
generate  some strong phases, which are important for estimating CP
violation.  However, unlike the vertex-type correction amplitude,
the calculation of annihilation amplitude  always suffers from
end-point divergence in collinear factorization approach. In the
pQCD approach, such divergence is regulated by the parton transverse
momentum $k_{T}$ at expense of modeling  additional $k_{T}$
dependence of meson distribution functions~\cite{KLS},  and a large
strong phase is found. In the QCD factorization (QCDF)
approach~\cite{Beneke1}, to give a conservative estimation, the
divergence is parameterized by complex parameters, $X_A=\int^{1}_{0}
dy/y=\mathrm{ln}(m_b/\Lambda) (1+\rho_A e^{i\phi_{A}})$, with
$\rho_A \leq 1$ and unrestricted $\phi_{A}$, which will sometimes
introduce large theoretical uncertainties in the final results. In
Refs.~\cite{PikSCET, Hurth}, annihilation diagram is studied with
SCET and also parameterized by a complex amplitude.  At present, the
dynamical origin of these corrections still remains a theoretical
challenge.

In this paper, we will revisit $B\to\pi K,~\pi K^{\ast}$ and $\rho
K$ decays within QCDF framework. However,  we shall quote the
infrared finite gluon propagator of Cornwall
prescription~\cite{Cornwall} to regulate these  divergences in
hard-sepctator scattering and annihilation amplitudes. With this
alternative scheme, we could evaluate both the strength and the
strong phase of hard spectator and annihilation corrections at the
expense of a dynamic gluon mass, which will be fitted in the twelve
decay modes.  It is interesting to note that the  infrared finite
behavior of gluon propagator are not only obtained from  solving the
well known Schwinger-Dyson equation~\cite{Cornwall,Aguilar,Alkofer},
but also supported by recent Lattice QCD simulations~\cite{lattice}.
Numerically, a sizable strength and a large strong phase of
annihilation corrections are found. Except $A_{CP}(B^{\pm}\to
K^{\pm}\pi^{0})$, our predictions for most of the branching ratios
and the direct CP asymmetries of $B\to\pi K$, $\pi K^{\ast}$ and
$\rho K$  agree with the current experimental data with an effective
gluon mass $m_g=0.45\sim0.55~{\rm GeV}$. However,  we get
$A_{CP}(B^{\pm}\to K^{\pm}\pi^{0})=-0.109\pm0.008$ which is still in
sharp contrast to experimental data $0.050\pm0.025$. To resolve this
mismatch,  we perform a model-independent analysis of new physics
contributions with   a set of  flavor-changing neutral current
(FCNC) $\bar{s}(1+\gamma_{5})b\otimes\bar{q}(1+\gamma_{5})q$ (q=u,d)
operators. To fit the twelve decay modes,  parameter spaces are
found generally with large weak phases. Our results indicate that
both strong phase from annihilation amplitude and new weak phase
from new physics are needed to account for the experimental data.

In Section~2, we revisit $B\to\pi K,~\pi K^{\ast}$ and $\rho K$
decays in the SM with QCDF modified by an infrared finite gluon
propagator for annihilation and spectator scattering kernels. After
recalculating the hard-spectator scattering  and the weak
annihilation corrections, we present our numerical results and
discussions. In Section~3, to find resolution to the CP violation
difference $\Delta A$, we present analyses of NP operators. Then, using the 
constrained parameters for the operators, we discuss the mixing-induced 
CP violations in $B\to \pi^{0}K_{S}$ and $\rho^{0}K_{S}$.  
Section~4 contains our conclusions. Appendix~A recapitulates the
decay amplitudes for the twelve decay modes within the
SM~\cite{Beneke3}. All the theoretical input parameters are
summarized in Appendix~B.

\section{ Revisiting $B\to\pi K, \pi K^{\ast}$ and $\rho K$ Decays in the SM}

In the SM, the effective weak Hamiltonian responsible for $b\to s$
transitions is given as~\cite{Buchalla:1996vs}
%%%%%%%%%%%%%%%%%%%%%%%%%%%%%%%%%%%%%%%%%%%%%%%%%
\begin{eqnarray}\label{eq:eff}
 {\cal H}_{\rm eff} &=& \frac{G_F}{\sqrt{2}} \biggl[V_{ub}
 V_{us}^* \left(C_1 O_1^u + C_2 O_2^u \right) + V_{cb} V_{cs}^* \left(C_1
 O_1^c + C_2 O_2^c \right) - V_{tb} V_{ts}^*\, \big(\sum_{i = 3}^{10}
 C_i O_i \big. \biggl. \nonumber\\
 && \biggl. \big. + C_{7\gamma} O_{7\gamma} + C_{8g} O_{8g}\big)\biggl] +
 {\rm h.c.},
\end{eqnarray}
%%%%%%%%%%%%%%%%%%%%%%%%%%%%%%%%%%%%%%%%%%%%%%%%%
where $V_{qb} V_{qs}^*$~($q=u, c$ and $t$) are products of the
Cabibbo-Kobayashi-Maskawa~(CKM) matrix elements~\cite{ckm}, $C_{i}$
the Wilson coefficients, and $O_i$ the relevant four-quark operators
whose explicit forms could be found, for example, in
Refs.~\cite{Beneke2, Buchalla:1996vs}.

In recent years, QCDF has been employed extensively to study the B
meson non-leptonic decays. For example, all of the decay modes
considered here have been studied comprehensively within the SM in
Refs.~\cite{Beneke2, Beneke3, YD, DSDu}. The relevant decay
amplitudes for $B\to\pi K$, $\pi K^{\ast}$ and $\rho K$ decays
within the QCDF formalism are shown in Appendix A. It is also noted
that the framework contains estimates of some power-suppressed but
numerically important contributions, such as the annihilation
corrections. However, due to the appearance of endpoint divergence,
these terms usually could not be computed rigorously. In
Refs.~\cite{Beneke2,Beneke3}, to probe their possible effects
conservatively,
 the endpoint divergent integrals are treated as signs of infrared sensitive contribution
 and  phenomenological parameterized by
%%%%%%%%%%%%%%%%%%%%%%%%%%%%%%%%%%%%%%%%%%%%%%%%%
\begin{equation}\label{treat-for-anni}
\int_0^1 \frac{\!dx}{x}\, \to X_A =(1+\rho_A e^{i\phi_A}) \ln
\frac{m_B}{\Lambda_h}, \qquad \int_0^1dy \frac{\textmd{ln}y}{y}\,
\to -\frac{1}{2}(X_A)^2
 \end{equation}
%%%%%%%%%%%%%%%%%%%%%%%%%%%%%%%%%%%%%%%%%%%%%%%%%
with $\rho_A \leq 1$ and $\phi_A$ unrestricted. The different
scenarios corresponding to different choices of $\rho_A$ and
$\phi_A$ have been thoroughly discussed in Ref.~\cite{Beneke3}.
Although this way of parametrization seems reasonable, it is still
very worthy to find some alternative schemes to regulate these
endpoint divergences, as precise as possible, to estimate the
strength and the associated strong phase in these power suppressed
contributions.

It is interesting to note that recent theoretical and
phenomenological studies are now accumulating supports for a softer
infrared behavior of the gluon propagator~\cite{Alkofer,theo,phe}.
Furthermore, an infrared finite dynamical gluon propagator, which is
shown to be not divergent as fast as $\frac{1}{q^2}$, has  been
successfully applied to the B meson non-leptonic
decays~\cite{YYgluon, Natale}.  Following these studies, in this
paper we adopt the gluon propagator derived by
Cornwall~\cite{Cornwall}, to regulate the endpoint divergent
integrals encountered within the QCDF formalism. The infrared finite
gluon propagator is given by~(in  Minkowski space)~\cite{Cornwall}
%%%%%%%%%%%%%%%%%%%%%%%%%%%%%%%%%%%%%%%%%%%%%%%%%
\begin{eqnarray}
D(q^2)=\frac{1}{q^2-M_g^2(q^2)+i\epsilon}~,
 \label{Dg}
\end{eqnarray}
%%%%%%%%%%%%%%%%%%%%%%%%%%%%%%%%%%%%%%%%%%%%%%%%%
where $q$ is the gluon momentum. The corresponding strong coupling
constant reads
%%%%%%%%%%%%%%%%%%%%%%%%%%%%%%%%%%%%%%%%%%%%%%%
\begin{eqnarray}
\alpha_s(q^2)=\frac{4\pi}{\beta_0\mathrm{ln}\Big(\frac{q^2+4M_g^2(q^2)}{\Lambda_{QCD}^2}\Big)}~,
\label{Alphas}
\end{eqnarray}
%%%%%%%%%%%%%%%%%%%%%%%%%%%%%%%%%%%%%%%%%%%%%%%%%
where $\beta_0=11-\frac{2}{3}n_f$ is the first coefficient of the
beta function, and $n_f$ the number of active flavors. The dynamical
gluon mass $M_g^2(q^2)$ is obtained as~\cite{Cornwall}
%%%%%%%%%%%%%%%%%%%%%%%%%%%%%%%%%%%%%%%%%%%%%%%%%
\begin{eqnarray}
M_g^2(q^2)=m_g^2\Bigg[\frac{\mathrm{ln}\Big(\frac{q^2+4m_g^2}{\Lambda_{QCD}^2}\Big)}
{\mathrm{ln}\Big(\frac{4m_g^2}{\Lambda_{QCD}^2}\Big)}\Bigg]^{-\frac{12}{11}},
\label{Mg}
\end{eqnarray}
%%%%%%%%%%%%%%%%%%%%%%%%%%%%%%%%%%%%%%%%%%%%%%%%%
where $m_g$ is the effective gluon mass, with a typical value
$m_g=500\pm200~{\rm MeV}$, and $\Lambda_{QCD}=225~{\rm MeV}$.
%%%%%%%%%%%%%%%%%%%%%%%%%%%%%%%%%%%%%%%%%%%%%%%%%%
\begin{figure}[t]
\epsfxsize=13cm \centerline{\epsffile{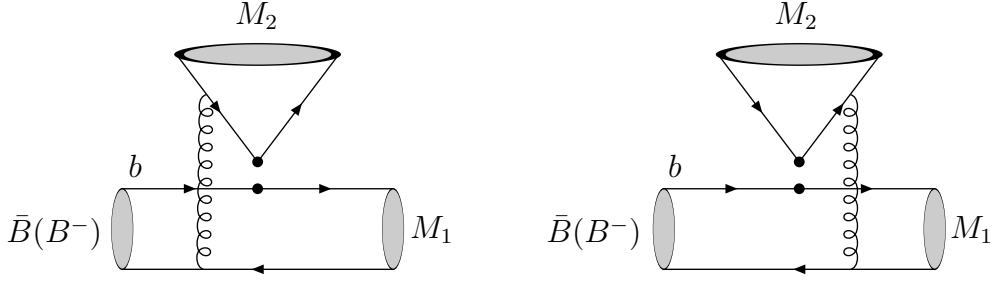}}
\centerline{\parbox{16cm}{\caption{\label{spec} \small Feynman
diagrams of hard spectator-scattering contributions.}}}
\end{figure}
%%%%%%%%%%%%%%%%%%%%%%%%%%%%%%%%%%%%%%%%%%%%%%%%%

\subsection{Recalculate the hard-spectator scattering and the annihilation contributions}
The next-to-leading order penguin contractions and vertex-type corrections to these decays
 are known free of infrared divergence and  well-defined in QCDF~\cite{Beneke2,Beneke3,YD},
for which  we would not repeat the calculation and concentrate on
the hard-spectator scattering and the annihilation contributions.
With the infrared finite gluon propagator to deal with the  endpoint
divergences, we will re-calculate the hard spectator and the
annihilation corrections in $B\to PP$ and $PV$ decays. The hard
spectator scattering Feynman diagrams are shown in Fig.~\ref{spec},
where the spectator anti-quark goes from the $\bar{B}$ meson to the
final-state $M_1$ meson and the $M_2$ meson is emitted from the weak
vertex. The longitudinal momentum fraction of the constituent quark
in the $M_{2(1)}$ meson is denoted by $x~(y)$, and $\xi$ is the
light-cone momentum fraction of the light anti-quark in the B meson.
To leading power in $1/m_b$, the hard spectator scattering
contributions can be expressed as~(where $x,\,y\gg\xi$ is assumed)
%%%%%%%%%%%%%%%%%%%%%%%%%%%%%%%%%%%%%%%%%%%%%%%%%
\begin{equation}
H_i(M_1M_2)= \frac{B_{M_1 M_2}}{A_{M_1 M_2}} \int_0^1dxdyd\xi
\frac{\alpha_s(q^2)}{\xi}\Phi_{B1}(\xi)\Phi_{M_2}(x)\Big[\frac{\Phi_{M_1}(y)}{\bar{x}(
\bar{y}+\omega^2(q^2)/\xi)}+r_\chi^{M_1}\frac{\phi_{m_1}(y)}{x(\bar{y}+\omega^2(q^2)/\xi)}\Big],
\label{hard1}
\end{equation}
%%%%%%%%%%%%%%%%%%%%%%%%%%%%%%%%%%%%%%%%%%%%%%%%%
for the contributions of operators $Q_{i=1-4,9,10}$,
%%%%%%%%%%%%%%%%%%%%%%%%%%%%%%%%%%%%%%%%%%%%%%%%%
\begin{equation}
H_i(M_1M_2)= -\frac{B_{M_1 M_2}}{A_{M_1 M_2}} \int_0^1dxdyd\xi
\frac{\alpha_s(q^2)}{\xi}\Phi_{B1}(\xi)\Phi_{M_2}(x)\Big[\frac{\Phi_{M_1}(y)}{x(
\bar{y}+\omega^2(q^2)/\xi)}+r_\chi^{M_1}\frac{\phi_{m_1}(y)}{\bar{x}(\bar{y}+\omega^2(q^2)/\xi)}\Big],
\label{hard2}
\end{equation}
%%%%%%%%%%%%%%%%%%%%%%%%%%%%%%%%%%%%%%%%%%%%%%%%%
for $Q_{i=5,7}$, and $H_i(M_1M_2)=0$ for $Q_{i=6,8}$.

In the above Eqs.~(\ref{hard1}) and (\ref{hard2}), $\Phi_{B1}(\xi)$
is the B meson light-cone distribution amplitude(LCDA),
$\Phi_{M_1}(x)$ and $\phi_{m_1}(y) $ are the twist-2 and the twist-3
LCDAs of light mesons, respectively, which are listed in Appendix B.
$\omega^2(q^2)=M_g^2(q^{2})/M_B^2$, $q^{2}=-Q^{2}$ and
$Q^2\simeq-\xi\bar{y}M_B^2$ is the space-like gluon momentum square
in the scattering kernels. The quantities $A_{M_{1}M_{2}}$ and
$B_{M_{1}M_{2}}$ collect relevant constants which can be found in
Ref.~\cite{Beneke3}.
%%%%%%%%%%%%%%%%%%%%%%%%%%%%%%%%%%%%%%%%%%%%%%%%%%
\begin{figure}[t]
\epsfxsize=13cm \centerline{\epsffile{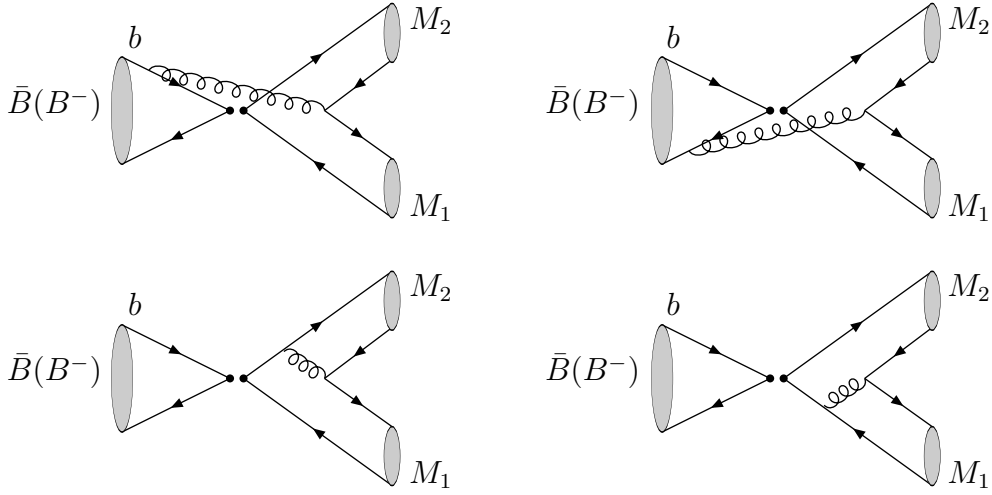}}
\centerline{\parbox{16cm}{\caption{\label{anni} \small Feynman
diagrams of weak annihilation contributions.}}}
\end{figure}
%%%%%%%%%%%%%%%%%%%%%%%%%%%%%%%%%%%%%%%%%%%%%%%

The Feynman diagrams of the weak annihilation topologies are shown
in Fig.~\ref{anni}. When both $M_1$ and $M_2$ are pseudoscalars, the
final decay amplitudes can be expressed as
%%%%%%%%%%%%%%%%%%%%%%%%%%%%%%%%%%%%%%%%%%%%%%%%%
\begin{eqnarray}
A_1^i&=&\pi\int_0^1dxdy\alpha_s(q^2)\biggl\{\Big[
\frac{\bar{x}}{(\bar{x}y-\omega^2(q^2)+i\epsilon)(1-x\bar{y})}+
\frac{1}{(\bar{x}y-\omega^2(q^2)+i\epsilon)\bar{x}}\Big]\Phi_{M_1}(y)\Phi_{M_2}(x)\nonumber\\
&&+\frac{2}{\bar{x}y-\omega^2(q^2)+i\epsilon}r_\chi^{M_1}r_\chi^{M_2}\phi_{m_1}(y)\phi_{m_2}(x)\biggl\}~,\label{anni1}\\
\nonumber\\
A_1^f&=&A_2^f=0,~\label{anni2}
\end{eqnarray}
%%%%%%%%%%%%%%%%%
\begin{eqnarray}
 A_2^i&=&\pi\int_0^1dxdy\alpha_s(q^2)\biggl\{\Big[
\frac{y}{(\bar{x}y-\omega^2(q^2)+i\epsilon)(1-x\bar{y})}+
\frac{1}{(\bar{x}y-\omega^2(q^2)+i\epsilon)y}\Big]\Phi_{M_1}(y)\Phi_{M_2}(x)\nonumber\\
&&+\frac{2}{\bar{x}y-\omega^2(q^2)+i\epsilon}r_\chi^{M_1}r_\chi^{M_2}\phi_{m_1}(y)\phi_{m_2}(x)\biggl\},
~\label{anni3}
\end{eqnarray}
\begin{eqnarray}
A_3^i&=&\pi\int_0^1dxdy\alpha_s(q^2)\biggl\{\frac{2\bar{y}}{(\bar{x}y-\omega^2(q^2)+i\epsilon)(1-x\bar{y})}
r_\chi^{M_1}\phi_{m_1}(y)\Phi_{M_2}(x)\nonumber\\
&&-\frac{2x}{(\bar{x}y-\omega^2(q^2)+i\epsilon)(1-x\bar{y})}r_\chi^{M_2}(x)\phi_{m_2}(x)\Phi_{M_1}(y)\biggl\}~,\label{anni4}\\
\nonumber\\
A_3^f&=&\pi\int_0^1dxdy\alpha_s(q^2)\biggl\{\frac{2(1+\bar{x})}{(\bar{x}y-\omega^2(q^2)+i\epsilon)\bar{x}}
r_\chi^{M_1}\phi_{m_1}(y)\Phi_{M_2}(x)\nonumber\\
&&+\frac{2(1+y)}{(\bar{x}y-\omega^2(q^2)+i\epsilon)y}r_\chi^{M_2}(x)\phi_{m_2}(x)\Phi_{M_1}(y)\biggl\}~,\label{anni5}
\end{eqnarray}
%%%%%%%%%%%%%%%%%%%%%%%%%%%%%%%%%%%%%%%%%%%%%%%%%
where $q^2\simeq\bar{x}yM_B^2$ is the time-like gluon momentum
square. The ``chirally-enhanced" factor $r_\chi^{M}$ is presented in
Appendix B. The superscript ``$i$" and ``$f$" refer to the gluon
emission from initial- and final-state quarks, respectively. The
subscript ``1", ``2", and ``3" correspond to three possible Dirac
structure, with ``1" for $(V-A)\otimes(V-A)$, ``2" for
$(V-A)\otimes(V+A)$, and ``3" for $(S-P)\otimes(S+P)$, respectively.
When $M_1$ is a vector meson and $M_2$ a pseudoscalar, the sign of
the second term in $A_1^i$, the first term in $A_2^i$, and the
second terms in $A_3^i$ and $A_3^f$  are needed to be changed. When
$M_2$ is a vector meson and $M_1$ a pseudoscalar, one only has to
change the overall sign of $A_2^i$.

%%%%%%%%%%%%%%%Fig.1%%%%%%%%%%%%%%%%%%%%%%%%%
\begin{figure}[h]
\begin{center}
\epsfxsize=15cm \centerline{\epsffile{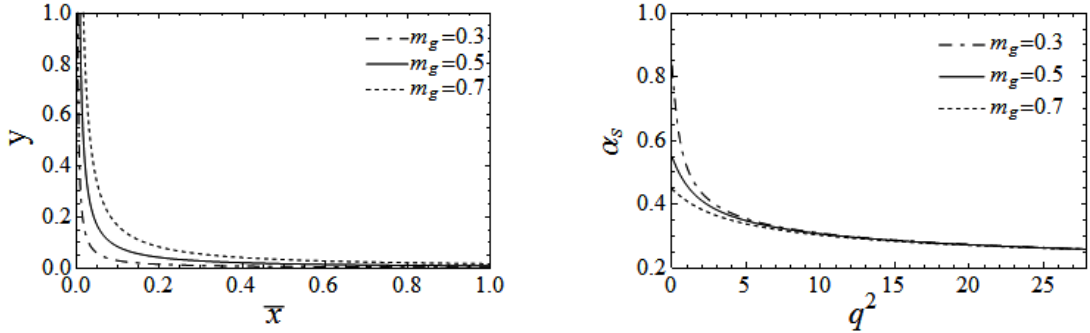}}
\centerline{\parbox{16cm}{\caption{\label{singularity_alphas}\small
The singularities in integral spaces~(left figure) in annihilation
contributions and the variations of strong coupling constant
corresponding to different $m_g$ choices~(in unit of ${\rm
GeV}$).}}}
\end{center}
\end{figure}
%%%%%%%%%%%%%%%%%%%%%%%%%%%%%%%%%%%%%%%%%%%
As shown by Eqs.~(\ref{hard1}) and  (\ref{hard2}) of the
hard-spectator scattering  contributions, the endpoint divergences
are regulated by the infrared finite form of the gluon propagator.
It is easy to observe from Eqs.~(\ref{hard1}) and (\ref{hard2}) that
hard-spectator scattering  contributions are real. For the
annihilation contributions shown by
Eqs.~(\ref{anni1})--(\ref{anni5}), singularities of the time-like
gluon propagators at the end-point of integrations (end-point
divergence) are moved into integral intervals with the infrared
finite form of the gluon propagator.  Singularities in the integral
intervals and variations of the effective strong coupling constant
are shown in Fig.~\ref{singularity_alphas}. It is noted  that
effective strong coupling constant is finite, but rather large in
the small $q^{2}$ region. However, there is strong cancellations
among the contributions  of the small $q^{2}$ region nearby
$m_{g}^{2}$, which renders  the annihilation contribution dominated
by  $q^{2}> m_{g}^{2}$  region associated  with a large imaginary
part. This situation is quite similar to pQCD~\cite{KLS} where the
large imaginary part from propagator  regulated by $k_{T}$
\begin{equation}
\frac{1}{xy m^{2}_{B}-k^{2}_{T}+i \epsilon}=P (\frac{1}{xy
m^{2}_{B}-k^{2}_{T}})-i \pi \delta(xy m^{2}_{B}-k^{2}_{T}),
\end{equation}
and it is also found the power suppression of these terms relative
to the leading  contributions was not very significant, and
important to account for CP violations in $B\to \pi K$ decays.

%%%%%%%%%%%%%Table 1%%%%%%%%%%%%%%%%%%%%%%%%%%%
\begin{table}[t]
 \begin{center}
 \caption{The $CP$-averaged branching ratios~(in units of $10^{-6}$)
 of  $B$ ${\to}$ $\pi K$, $\pi K^{\ast}$  and $\rho K$ decays in SM with
 different $m_g$ (in unit of ${\rm GeV}$) are presented in QCDF columns.}
 \label{tab_br}
 \vspace{0.5cm}
 \doublerulesep 0.7pt \tabcolsep 0.07in
 \begin{tabular}{lccccccccccc} \hline \hline
 \multicolumn{1}{c}{Decay Mode}&\multicolumn{3}{c}{QCDF}&  \multicolumn{1}{c}{Experiment} \\
 & $m_g=0.3$ & $m_g=0.7$ & $m_g=0.45\sim0.55$ &\multicolumn{1}{c}{data}\\ \hline
 $B_{u}^{-}$ ${\to}$ ${\pi}^{-}{\overline{K}}^{0}$ & $44.4$ & $16.8$ & $23.17\pm3.28$  &$23.1\pm1.0$\\
 $B_{u}^{-}$ ${\to}$ ${\pi}^{0}K^{-}$ &$23.4$ & $9.3$ & $12.50\pm1.65$ & $12.9\pm0.6$\\
 ${\overline{B}}_{d}^{0}$ ${\to}$ ${\pi}^{+}K^{-}$ & $44.7$ & $16.3$ & $22.71\pm3.27$ & $19.4\pm0.6$\\
 ${\overline{B}}_{d}^{0}$ ${\to}$ ${\pi}^{0}{\overline{K}}^{0}$ &$21.2$ & $7.3$ & $10.50\pm1.63$ & $9.9\pm0.6$\\
  \hline
 $B_{u}^{-}$ ${\to}$ ${\pi}^{-}{\overline{K}}^{{\ast}0}$ & $28.3$ & $5.2$ & $8.90\pm1.59$ &  $10.0\pm0.8$\\
 $B_{u}^{-}$ ${\to}$ ${\pi}^{0}K^{{\ast}-}$ &$15.2$ & $3.4$ & $5.25\pm0.83$ & $6.9\pm2.3$\\
 ${\overline{B}}_{d}^{0}$ ${\to}$ ${\pi}^{+}K^{{\ast}-}$ &$28.7$ & $5.3$ & $9.13\pm1.68$ &  $10.6\pm 0.9$\\
 ${\overline{B}}_{d}^{0}$ ${\to}$ ${\pi}^{0}{\overline{K}}^{{\ast}0}$ &$13.4$ & $1.9$ & $3.89\pm0.82$ & $2.4\pm0.7$\\
  \hline
 $B_{u}^{-}$ ${\to}$ ${\rho}^{-}{\overline{K}}^{0}$ &$31.8$ & $5.6$ & $10.27\pm1.96$ & $8.0^{+1.5}_{-1.4 }$\\
 $B_{u}^{-}$ ${\to}$ ${\rho}^{0}K^{-}$ &$14.9$ & $2.5$ & $4.81\pm0.94$ & $3.81^{+0.48}_{-0.46 }$\\
 ${\overline{B}}_{d}^{0}$ ${\to}$ ${\rho}^{+}K^{-}$ &$38.6$ & $8.0$ & $13.42\pm2.31$ & $8.6^{+0.9}_{-1.1 }$\\
 ${\overline{B}}_{d}^{0}$ ${\to}$ ${\rho}^{0}{\overline{K}}^{0}$ &$21.0$ & $4.8$ & $7.53\pm1.25$ & $5.4^{+0.9}_{-1.0 }$\\
 \hline \hline
 \end{tabular}
 \end{center}
 \end{table}
%%%%%%%%%%%%%%%%%%%%%%%%%%%%%%%%%%%%%%%%%%%%

%%%%%%%%%%%%%%%%Table 2%%%%%%%%%%%%%%%%%%%%%%%%
\begin{table}[t]
 \begin{center}
 \caption{The direct CP asymmetries ( in unit of $10^{-2}$) of $B$
 ${\to}$ $\pi K$, $\pi K^{\ast}$ and $\rho K$ decays in SM with different
 $m_g$ ~(in unit of ${\rm GeV}$). Other captions are the same as Table 1.}
 \label{tab_cp}
 \vspace{0.5cm}
 \doublerulesep 0.7pt \tabcolsep 0.07in
 \begin{tabular}{lccccccccccc} \hline \hline
 \multicolumn{1}{c}{Decay Mode}&\multicolumn{3}{c}{QCDF}&\multicolumn{1}{c}{Experiment} \\
 & $m_g=0.3$ & $m_g=0.7$ & $m_g=0.45\sim0.55$ & \multicolumn{1}{c}{data}\\ \hline
 $B_{u}^{-}$ ${\to}$ ${\pi}^{-}{\overline{K}}^{0}$ & $0.06$ & $0.19$ & $0.10\pm0.08$ &$0.9\pm2.5$\\
 $B_{u}^{-}$ ${\to}$ ${\pi}^{0}K^{-}$ &$-11.6$ & $-8.3$ & $-10.85\pm0.84$ & $5.0\pm2.5$\\
 ${\overline{B}}_{d}^{0}$ ${\to}$ ${\pi}^{+}K^{-}$ & $-11.0$ & $-11.4$ & $-12.38\pm0.69$ & $-9.7\pm1.2$\\
 ${\overline{B}}_{d}^{0}$ ${\to}$ ${\pi}^{0}{\overline{K}}^{0}$ &$2.5$ & $0.1$ & $1.39\pm0.35$ & $-14\pm11$\\
  \hline
 $B_{u}^{-}$ ${\to}$ ${\pi}^{-}{\overline{K}}^{{\ast}0}$ & $0.3$ & $-0.0$ & $0.16\pm0.16$ & $-11.4\pm6.1$\\
 $B_{u}^{-}$ ${\to}$ ${\pi}^{0}K^{{\ast}-}$ &$-27.0$ & $-34.1$ & $-41.20\pm6.69$ & $4\pm29$\\
 ${\overline{B}}_{d}^{0}$ ${\to}$ ${\pi}^{+}K^{{\ast}-}$ &$-27.2$ & $-47.6$ & $-47.58\pm8.42$ & $-10\pm11$\\
 ${\overline{B}}_{d}^{0}$ ${\to}$ ${\pi}^{0}{\overline{K}}^{{\ast}0}$ &$3.9$ & $2.1$ & $4.67\pm1.14$ & $-9^{+32}_{-23}$\\
  \hline
 $B_{u}^{-}$ ${\to}$ ${\rho}^{-}{\overline{K}}^{0}$ &$0.1$ & $1.2$ & $0.53\pm0.21$ & $-12\pm17$\\
 $B_{u}^{-}$ ${\to}$ ${\rho}^{0}K^{-}$ &$28.1$ & $49.7$ & $46.27\pm5.94$ & $37\pm11 $\\
 ${\overline{B}}_{d}^{0}$ ${\to}$ ${\rho}^{+}K^{-}$ &$19.3$ & $31.5$ & $31.40\pm4.63$ & $15\pm13$\\
 ${\overline{B}}_{d}^{0}$ ${\to}$ ${\rho}^{0}{\overline{K}}^{0}$ &$-4.2$ & $0.2$ & $-3.26\pm1.29$ & $-2\pm29$\\
 \hline \hline
 \end{tabular}
 \end{center}
 \end{table}
%%%%%%%%%%%%%%%%%%%%%%%%%%%%%%%%%%%%%%%%%%%%

\subsection{The branching ratios and direct CP asymmetries in the SM}
With the prescriptions for the endpoint divergences, we will present
our numerical results of branching ratios and  CP violations in
these decays. Decay amplitudes and input parameters are listed in
Appendices A and B, respectively. Our results are summarized in
Table~\ref{tab_br} and Table~\ref{tab_cp}, where the relevant
experimental data  are also tabled for comparison.

In Table~\ref{tab_br} (\ref{tab_cp}), the experimental data column
is the up-to-date averages for these branching ratios (direct CP
violations) by HFAG~\cite{HFAG}. It is shown that all the results
are in good agreements with the experimental data with
$m_g=0.45\sim0.55~{\rm GeV}$. It is also noted  that the dynamical
gluon mass $m_g=0.45\sim0.55~{\rm GeV}$ are also  consistent with
findings in other phenomenal  studies of B decays~\cite{YYgluon,
Natale} and the different solutions of SDE~\cite{Cornwall, Aguilar,
Alkofer}. The phenomenology  successes may indicate that the gluon
mass, although not a directly measurable quantity, furnishes a
regulator for infrared divergences of QCD scattering processes.

From the CP averaged branching ratios in the fourth column of
Table~\ref{tab_br}, we get
%%%%%%%%%%%%%%%%%%%%%%%%%%%%%%%%%%%%%%%%%%%%%%%%%
\begin{eqnarray}
&&R_c\equiv~2\bigg[\frac{Br(B^-\to\pi^0 K^-)}{Br(B^-\to\pi^-
K^0)}\bigg]=1.08\pm0.30 ,\nonumber\\
&&R_n\equiv~\frac{1}{2}\bigg[\frac{Br(\bar{B}^0\to\pi^+
K^-)}{Br(\bar{B}^0\to\pi^0
K^0)}\bigg]=1.08\pm0.32,
\end{eqnarray}
%%%%%%%%%%%%%%%%%%%%%%%%%%%%%%%%%%%%%%%%%%%%%%%%%
which agree with the experimental data $R_c=1.12\pm0.10$ and
$R_n=0.98\pm0.09$~\cite{HFAG}.
%%%%%%%%%%%%%%%%%%%

Table~\ref{tab_cp} is our results for direct CP violations. The
fourth column  is the results estimated with
$m_g=0.45\sim0.55~{\rm GeV}$ fixed by branching ratios, where the
error-bars are simply due to the $m_{g}$ variations. Compared with
the experimental data, our results, except
$A_{CP}(B^{-}_u\to\pi^{0}K^{-})$,  agree with the measurements. For
the most significant experimental result  among the measurements of
direct CP violations in the twelve decay modes
 $A_{CP}(\bar{B}^0_d\to\pi^{+}K^{-})=-0.097\pm0.012$~\cite{HFAG}, our
result  $A_{CP}(\bar{B}^0_d\to\pi^{+} K^{-})=-0.124\pm0.007$ is in
 good agreement with it. As expected in the SM, we find
again $A_{CP}(B^{-}_u\to\pi^{0}K^{-})=-0.108\pm0.008$  very close to
$A_{CP}(\bar{B}^0_d\to\pi^{+} K^{-})$, which  are generally in
agreement with the results of Refs.~\cite{Beneke3, PikPQCD, PikSCET}
listed in Eq.~(\ref{AcpPuzzleQCD})--(\ref{AcpPuzzleSCET}). So, it is
very hard to accommodate the measured large difference between
$A_{CP}(B^{-}_u\to\pi^{0}K^{-})$ and $A_{CP}(\bar{B}^0_d\to\pi^{+}
K^{-})$ in the SM with the available approaches for hadron-dynamics
in B decays.

Although the problem could  be due to hadronic effects unknown so
far, the difference  between $A_{CP}(B^{-}_u\to\pi^{0}K^{-})$  and
$A_{CP}(\bar{B}^0_d\to\pi^{+} K^{-})$ could  be an indication of new
sources of CP violation beyond the SM~\cite{mannel,Baek,hou}.
%%%%%%%%%%%

\section{ Possible resolution with new  $(S+P)\otimes(S+P)$ operators }
In this  Section we will pursue possible NP solutions
model-independently with a set of FCNC $(S+P)\otimes(S+P)$
operators. The effects of anomalous tensor and (pseudo-)scalar
operators on hadronic B decays have attracted  many attentions
recently~\cite{Baek,kagan, KC,KCrecently, YYtensor,kundu2}. For
example, it is  shown that they could help to resolve the abnormally
large transverse polarizations observed in $B\to \phi K^\ast$ decay,
as well as the large $Br(B\to\eta K^\ast)$~\cite{YYtensor}.

The general four-quark tensor operators can be expressed as
%%%%%%%%%%%%%%%%%%%%%%%%%%%%%%%%%%%%%%%%%%%%%%%%%
\begin{eqnarray}\label{eq:tensor}
O^q_{T}&=&\bar{s}\sigma_{\mu\nu}(1+\gamma_{5})b\otimes\bar{q}\sigma^{\mu\nu}(1+\gamma_{5})q~,
\quad
 O^{\prime
q}_{T}=\bar{s}_{i}\sigma_{\mu\nu}(1+\gamma_{5})b_{j}\otimes\bar{q}_{j}\sigma^{\mu\nu}(1+\gamma_{5})q_{i}\,,
\end{eqnarray}
%%%%%%%%%%%%%%%%%%%%%%%%%%%%%%%%%%%%%%%%%%%%%%%%%
which could be expressed, through the Fierz transformations, as
linear combinations of the (pseudo-)scalar operators. In our present
case, however, we find that the tensor operators with q=u,d give the
same contributions to the $B^{-}_u\to\pi^{0}K^{-}$ and
$B^{0}_d\to\pi^{+}K^{-}$ decays, so that they are hardly possible to
resolve the direct CP violation difference, because after Fierz
transformations, $O^q_{T}$ and $O^{\prime q}_{T}$ with $q=u, d$ will
give operators like
$\bar{q}(1+\gamma_{5})b\otimes\bar{s}(1+\gamma_{5})q$ which are
different from $\bar{s}(1+\gamma_{5})b\otimes\bar{s}(1+\gamma_{5})s$
of the Fierz transforming
$O^s_{T}=\bar{s}\sigma_{\mu\nu}(1+\gamma_{5})b\otimes\bar{s}\sigma^{\mu\nu}(1+\gamma_{5})s$
for $B\to \phi K^\ast$ decays. On the other hand, the new operators
like $\bar{s}(1+\gamma_{5})b\otimes\bar{q}(1+\gamma_{5})q$ may give
a possible solution to $\Delta A$ because of their different
contributions to the $B^{-}\to\pi^{0}K^{-}$ and
$\bar{B}^{0}\to\pi^{+}K^{-}$ decays.

We write the NP effective Hamiltonian for $b\to s$ transitions as
%%%%%%%%%%%%%%%%%%%%%%%%%%%%%%%%%%%%%%%%%%%%%%%%%
\begin{equation}\label{eq:np}
{\cal H}_{eff}^{\rm NP}=\frac{G_{F}}{\sqrt{2}}\,\sum_{q=u,d}
|V_{tb}V_{ts}^{\ast}|{ e^{i\delta_{S}^{q}}\,\Big[C_{S1}^{q}O^q_{S1}
+C_{S8}^{q}O^q_{S8}\Big]}+ {\rm h.c.}~,
\end{equation}
%%%%%%%%%%%%%%%%%%%%%%%%%%%%%%%%%%%%%%%%%%%%%%%%%
with $O^q_{S1}$ and $O^q_{S8}$ defined by
%%%%%%%%%%%%%%%%%%%%%%%%%%%%%%%%%%%%%%%%%%%%%%%%%
\begin{eqnarray}\label{eq:sp}
O^u_{S1}&=&\bar{s}(1+\gamma_{5})b\otimes\bar{u}(1+\gamma_{5})u~,
\qquad
O^{u}_{S8}=\bar{s}_{i}(1+\gamma_{5})b_{j}\otimes\bar{u}_{j}(1+\gamma_{5})u_{i}~,\nonumber\\
O^d_{S1}&=&\bar{s}(1+\gamma_{5})b\otimes\bar{d}(1+\gamma_{5})d~,
\qquad O^{
d}_{S8}=\bar{s}_{i}(1+\gamma_{5})b_{j}\otimes\bar{d}_{j}(1+\gamma_{5})d_{i},
\end{eqnarray}
%%%%%%%%%%%%%%%%%%%%%%%%%%%%%%%%%%%%%%%%%%%%%%%%%
where $i$ and $j$ are color indices.  The
coefficient $C^q_{S1(S8)}$ describes the relative interaction
strength of the  operator $O^q_{S1(S8)}$, and
$\delta^q_{S}$ is their possible NP weak phase. Since both the coefficients and
the weak phase are unknown parameters, for simplicity, we shall only
consider their leading contributions with the naive factorization(NF) approximation.
%%%%%%%%%%%%%%%%%%%%%%%%%%%%%%%%%%%%%%%%
\begin{figure}[ht]
\begin{center}
\epsfxsize=15cm \centerline{\epsffile{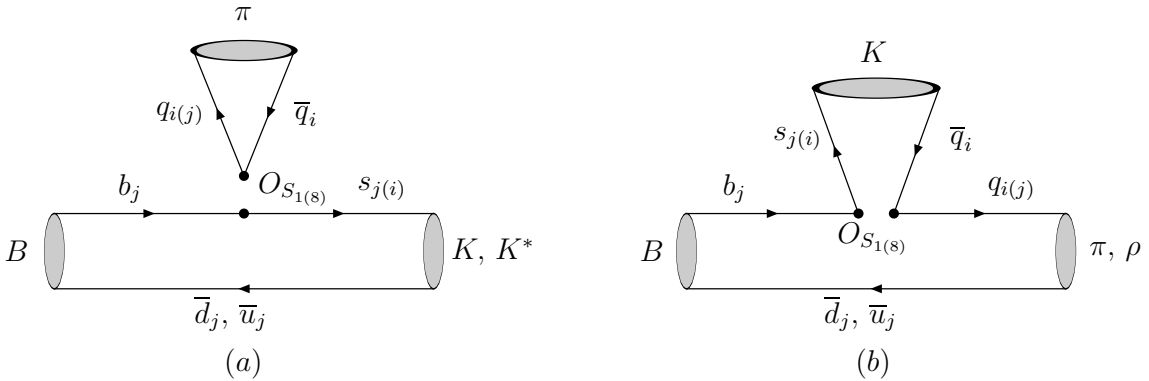}}
\centerline{\parbox{16cm}{\caption{\label{NPFeynman}\small Feynman
diagrams contributing to the amplitudes of $B\to\pi K, \pi K^{\ast}$
and $\rho K$ decays due to the  $(S+P)\otimes (S+P)$
operators.}}}
\end{center}
\end{figure}
%%%%%%%%%%%%%%%%%%%%%%%%%%%%%%%%%%%%%%%%%%%

The relevant Feynman diagrams of the NP operators are shown in
Fig.~\ref{NPFeynman} with $q=u,\,d$. With the NF approximation, it
is easy to see that, for the $B\to\pi^0 K^{\ast-}$ and
$\pi^0\bar{K}^{\ast0}$ decay modes, only Fig.~\ref{NPFeynman}~(a)
contributes, for the $B\to\pi^-\bar{K}^0$, $\pi^+K^-$  and $\rho K$
decay modes, only Fig.~\ref{NPFeynman}~(b) contributes, while both
topology structures  contribute to the $B\to\pi^0 K^{-}$ and
$\pi^0\bar{K}^{0}$ decay modes. However,
 none of them contributes to $B\to\pi^- K^{\ast0}$ and $\pi^+
K^{\ast-}$ decays.  After some simple calculations, these NP
contributions to the decay amplitudes of the $B\to\pi K$, $\pi
K^{\ast}$  and $\rho K$ decays are obtained as
%%%%%%%%%%%%%%%%%%%%%%%%%%%%%%%%%%%%%%%%%%%%%%%%%
\begin{eqnarray}\label{eq:NPcontribution}
{\cal A}^{\rm NP}_{B^{-}\rightarrow\pi^-
\bar{K}^{0}}&=&i\,\frac{G_{F}}{\sqrt{2}}\,\frac{1}{4}\,
|V_{tb}V_{ts}^{\ast}|\, m_{B_u}^2\, e^{i\delta_{S}^d}\,g_{S}^d \,
r_{\chi}^K\, F_0^{B\rightarrow \pi}(m_K^2)\,f_{K}~,
\label{amp1_NP}\\
{\cal A}^{\rm NP}_{B^{-}\rightarrow\pi^0
K^{-}}&=&i\,\frac{G_{F}}{\sqrt{2}}\,\frac{1}{4\sqrt{2}}\,
|V_{tb}V_{ts}^{\ast}|\, m_{B_u}^2\,\Big[ e^{i\delta_{S}^u}\,g_{S}^u
\,r_{\chi}^K\,F_0^{B\rightarrow
\pi}(m_K^2)\,f_{K}\,\nonumber\\
&&-\,2\,\big(e^{i\delta_{S}^u}\,g_{S}^{\prime
u}-e^{i\delta_{S}^d}\,g_{S}^{\prime d}\big)
\,r_{\chi}^{\pi}\,F_0^{B\rightarrow K}(m_\pi^2)\,f_{\pi}\Big]~,\label{amp2_NP}\\
{\cal A}^{\rm NP}_{\bar{B}^{0}\rightarrow\pi^+
K^-}&=&i\,\frac{G_{F}}{\sqrt{2}}\,\frac{1}{4}\,
|V_{tb}V_{ts}^{\ast}|\, m_{B_d}^2\, e^{i\delta_{S}^u}\,g_{S}^u \,
r_{\chi}^K\, F_0^{B\rightarrow \pi}(m_K^2)\,f_{K}~,\label{amp3_NP}\\
{\cal A}^{\rm NP}_{\bar{B}^{0}\rightarrow\pi^0
\bar{K}^0}&=&i\,\frac{G_{F}}{\sqrt{2}}\,\frac{1}{4\sqrt{2}}\,
|V_{tb}V_{ts}^{\ast}|\, m_{B_d}^2\,\Big[-
e^{i\delta_{S}^d}\,g_{S}^d\,
r_{\chi}^K\, F_0^{B\rightarrow \pi}(m_K^2)\,f_{K}\nonumber\\
&&-\,2\big( e^{i\delta_{S}^u}\,g_{S}^{\prime
u}\,-\,e^{i\delta_{S}^d}\,g_{S}^{\prime d}\big)\,r_{\chi}^{\pi}\,
F_0^{B\rightarrow K}(m_{\pi}^2)\,f_{\pi}\Big]~,\label{amp4_NP}
\end{eqnarray}
\begin{eqnarray}
{\cal A}^{\rm NP}_{B^{-}\rightarrow\pi^- \bar{K}^{\ast 0}}&=&\,0~,\label{amp5_NP}\\
{\cal A}^{\rm NP}_{B^{-}\rightarrow\pi^0
K^{\ast-}}&=&\,i\,\frac{G_{F}}{\sqrt{2}}\,\frac{1}{2\sqrt{2}}\,
|V_{tb}V_{ts}^{\ast}|\,
m_{B_u}^2\,\Big[e^{i\delta_{S}^u}\,g_{S}^{\prime
u}\,-\,e^{i\delta_{S}^d}\,g_{S}^{\prime
d}\Big]\,r_{\chi}^{\pi}\,A_0^{B\rightarrow K^{\ast}}(m_{\pi}^2)\,f_{\pi},\label{amp6_NP}\\
{\cal A}^{\rm NP}_{\bar{B}^{0}\rightarrow\pi^+ K^{\ast -}}&=&\,0~,\label{amp7_NP}\\
{\cal A}^{\rm NP}_{\bar{B}^{0}\rightarrow\pi^0
\bar{K}^{\ast0}}&=&\,i\,\frac{G_{F}}{\sqrt{2}}\,\frac{1}{2\sqrt{2}}\,
|V_{tb}V_{ts}^{\ast}|\,
m_{B_u}^2\,\Big[e^{i\delta_{S}^u}\,g_{S}^{\prime
u}\,-\,e^{i\delta_{S}^d}\,g_{S}^{\prime
d}\Big]\,r_{\chi}^{\pi}\,A_0^{B\rightarrow
K^{\ast}}(m_{\pi}^2)\,f_{\pi},\label{amp8_NP}\\
%\end{eqnarray}
%\begin{eqnarray}
{\cal A}^{\rm NP}_{B^{-}\rightarrow\rho^-
\bar{K}^{0}}&=&\,-\,i\,\frac{G_{F}}{\sqrt{2}}\,\frac{1}{4}\,
|V_{tb}V_{ts}^{\ast}|\, m_{B_u}^2\,e^{i\delta_{S}^d}\,g_{S}^{
d}\,r_{\chi}^{K}\,A_0^{B\rightarrow \rho}(m_{K}^2)\,f_{K},\label{amp9_NP}\\
{\cal A}^{\rm NP}_{B^{-}\rightarrow\rho^0
K^{-}}&=&\,-\,i\,\frac{G_{F}}{\sqrt{2}}\,\frac{1}{4\sqrt{2}}\,
|V_{tb}V_{ts}^{\ast}|\, m_{B_u}^2\,e^{i\delta_{S}^u}\,g_{S}^{
u}\,r_{\chi}^{K}\,A_0^{B\rightarrow \rho}(m_{K}^2)\,f_{K},\label{amp10_NP}\\
{\cal A}^{\rm NP}_{\bar{B}^{0}\rightarrow\rho^+
K^{-}}&=&\,-\,i\,\frac{G_{F}}{\sqrt{2}}\,\frac{1}{4}\,
|V_{tb}V_{ts}^{\ast}|\, m_{B_u}^2\,e^{i\delta_{S}^u}\,g_{S}^{
u}\,r_{\chi}^{K}\,A_0^{B\rightarrow \rho}(m_{K}^2)\,f_{K},\label{amp11_NP}\\
{\cal A}^{\rm NP}_{\bar{B}^{0}\rightarrow\rho^0
\bar{K}^{0}}&=&\,i\,\frac{G_{F}}{\sqrt{2}}\,\frac{1}{4\sqrt{2}}\,
|V_{tb}V_{ts}^{\ast}|\,
m_{B_u}^2\,e^{i\delta_{S}^d}\,g_{S}^{d}\,r_{\chi}^{K}\,A_0^{B\rightarrow
\rho}(m_{K}^2)\,f_{K},\label{amp12_NP}
\end{eqnarray}
%%%%%%%%%%%%%%%%%%%%%%%%%%%%%%%%%%%%%%%%%%%%%%%%%
where
%%%%%%%%%%%%%%%%%%%%%%%%%%%%%%%%%%%%%%%%%%%%%%%%%
\begin{eqnarray}\label{NPPara_relation}
g_{S}^{\prime u}&=&\,C_{S1}^{u}\,+\,\frac{1}{N_c}\,C_{S8}^{u}\,,
~~~~~~g_{S}^{u}=\,C_{S8}^{u}\,+\,\frac{1}{N_c}\,C_{S1}^{u}\,,\nonumber\\
g_{S}^{\prime d}&=&\,C_{S1}^{d}\,+\,\frac{1}{N_c}\,C_{S8}^{d}\,,
~~~~~~g_{S}^{d}=\,C_{S8}^{d}\,+\,\frac{1}{N_c}\,C_{S1}^{d}\,.
\end{eqnarray}
%%%%%%%%%%%%%%%%%%%%%%%%%%%%%%%%%%%%%%%%%%%%%%%%%
Comparing the NP amplitudes Eq.~(\ref{amp2_NP}) with
Eq.~(\ref{amp3_NP}), we expect that these new (pseudo-)scalar
operators might provide a possible resolution to the direct CP violation difference,
which is  realized in the following numerical analyses.

\subsection{Numerical analyses and discussions of new pseudo-scalar operators}
Our analysis consists of five cases with different assumptions for
dominance of NP operators, namely,
\begin{center}
\begin{itemize}
\item Case I:   $b\to s u{\bar u}$ operators $O^u_{S1}$ and $O^u_{S8}$,
\item Case II:  $b\to s d{\bar d}$ operators $O^d_{S1}$ and $O^d_{S8}$,
\item Case III: $b\to s d{\bar d}$ operator $O^d_{S1}$ solely,
\item Case IV:  only color singlet operators $O^u_{S1}$ and $O^d_{S1}$,
\item Case V:  all the operators   $O^u_{S1}$, $O^u_{S8}$, $O^d_{S1}$ and $O^d_{S8}$.
\end{itemize}
\end{center}
For each case, the corresponding effective Hamiltonian could be read
from Eq.~(\ref{eq:np}). It could be  expected that  a collection of
related  decay modes could constrain the relevant  NP parameter
spaces restrictively.

\begin{figure}[ht]
\begin{center}
\epsfxsize=15cm \centerline{\epsffile{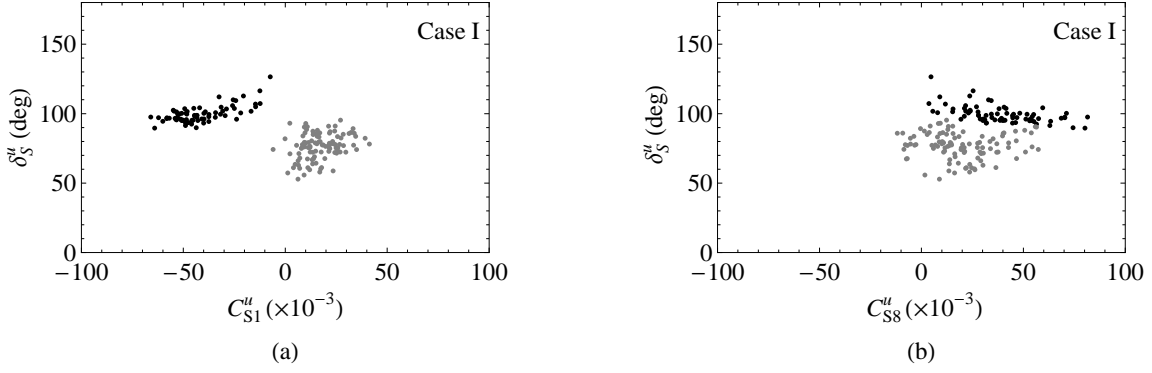}}
\centerline{\parbox{16cm}{\caption{\label{CaseI}\small The allowed
regions for the parameters $C_{S1}^{u}$, $C_{S8}^{u}$ and
$\delta_{S}^{u}$  of  Case I.}}}
\end{center}
\end{figure}

\begin{figure}[t]
\begin{center}
\epsfxsize=15cm \centerline{\epsffile{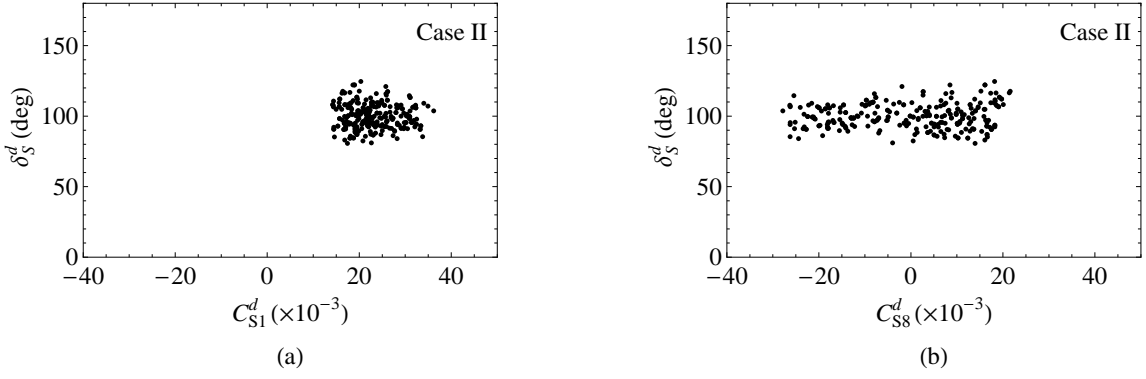}}
\centerline{\parbox{16cm}{\caption{\label{CaseII}\small The allowed
regions for the parameters $C_{S1}^{d}$, $C_{S8}^{d}$ and
$\delta_{S}^{d}$ in Case II with $m_g=0.5\,{\rm GeV}$.}}}
\end{center}
\end{figure}

\begin{figure}[ht]
\begin{center}
\epsfxsize=7.5cm \centerline{\epsffile{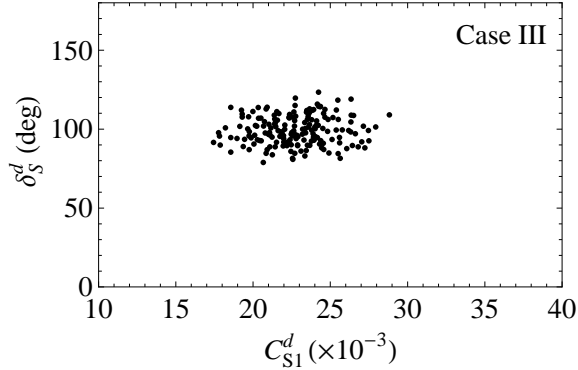}}
\centerline{\parbox{16cm}{\caption{\label{CaseIII}\small The allowed
regions for the parameters $C_{S1}^{d}$ and $\delta_{S}^{d}$  of  Case
III.}}}
\end{center}
\end{figure}

\begin{figure}[ht]
\begin{center}
\epsfxsize=15cm \centerline{\epsffile{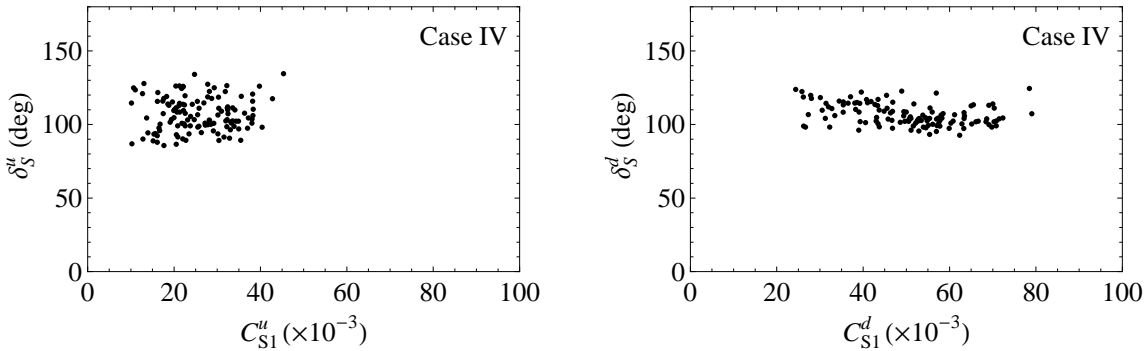}}
\centerline{\parbox{16cm}{\caption{\label{CaseIV}\small The allowed
regions for the parameters $C_{S1}^{u}$, $\delta_{S}^{u}$,
$C_{S1}^{d}$ and $\delta_{S}^{d}$ of  Case IV.}}}
\end{center}
\end{figure}

\begin{figure}[ht]
\begin{center}
\epsfxsize=15cm \centerline{\epsffile{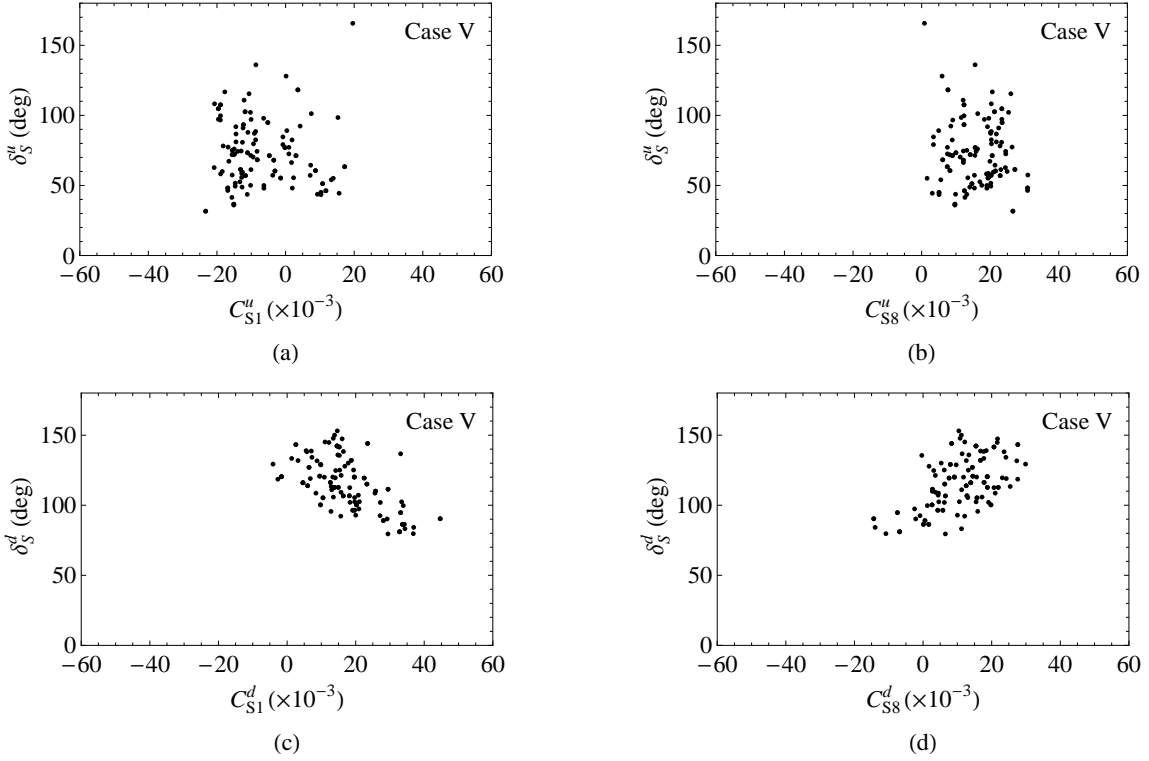}}
\centerline{\parbox{16cm}{\caption{\label{CaseV}\small The allowed
regions for the parameters $C_{S1}^{u}$, $C_{S8}^{u}$
$\delta_{S}^{u}$, $C_{S1}^{d}$, $C_{S8}^{d}$ and $\delta_{S}^{d}$ of
Case V.}}}
\end{center}
\end{figure}

Our fitting is performed with the experimental data varying randomly
within their $2\sigma$ error-bars, while the theoretical
uncertainties are obtained by varying the input  parameters within
the regions specified in Appendix~B. Our numerical results are
summarized in Table~\ref{tabNPbr}--\ref{NPSpace} where the assigned
uncertainties of our fitting results should be understood at
$2\sigma$ statistical level. Illustratively, the constrained  NP
parameter spaces are shown in Figs.~\ref{CaseI}--\ref{CaseV},
respectively. It is noted  that, to leading order approximation,
both $B_{u}^{-}\to \pi^- \bar{K}^{\ast 0}$ and $\bar{B}_{d}^{0}\to
\pi^+ \bar{K}^{\ast -}$ decays do not receive these NP
contributions, so we perform fitting for the remained ten decay
modes. In the following, we present numerical analyses subdivided into five cases. 
\par
%%%%%%%%%%%%%%%%%%%%%%%%%%%%%%%%%%%%%%%%
\begin{table}[t]
 \begin{center}
 \caption{The $CP$-averaged branching ratios~(in units of $10^{-6}$)  in different NP Cases with $m_g=0.5{\rm GeV}$. The  dash means (pseudo-)scalar operators of the Case irrelevant to the corresponding decay mode.}
 \label{tabNPbr}
 \vspace{0.5cm}
 \doublerulesep 0.7pt \tabcolsep 0.07in
 \begin{tabular}{lccccccccccc} \hline \hline
 \multicolumn{1}{c}{Decay Mode}&  \multicolumn{1}{c}{Experiment} & \multicolumn{5}{c}{NP}\\
 &\multicolumn{1}{c}{data} & \multicolumn{1}{c}{Case I} & \multicolumn{1}{c}{Case II}& \multicolumn{1}{c}{Case III}& \multicolumn{1}{c}{Case IV}& \multicolumn{1}{c}{Case V}\\ \hline
 $B_{u}^{-}$ ${\to}$ ${\pi}^{-}{\overline{K}}^{0}$ & $23.1\pm1.0$ & --- & $23.0\pm1.0$ & $22.9\pm0.9$ & $21.5\pm0.3$ & $22.4\pm0.9$\\
 $B_{u}^{-}$ ${\to}$ ${\pi}^{0}K^{-}$ & $12.9\pm0.6$ & $12.1\pm0.4$ & $12.8\pm0.7$ & $12.7\pm0.6$ & $12.1\pm0.3$ & $12.1\pm0.4$\\
 ${\overline{B}}_{d}^{0}$ ${\to}$ ${\pi}^{+}K^{-}$ & $19.4\pm0.6$& $20.2\pm0.3$ & --- & --- & $20.4\pm0.2$ & $20.1\pm0.4$\\
 ${\overline{B}}_{d}^{0}$ ${\to}$ ${\pi}^{0}{\overline{K}}^{0}$ & $9.9\pm0.6$ & $9.0\pm0.3$ & $9.9\pm0.6$ & $10.0\pm0.7$ &$9.0\pm0.2$ & $9.1\pm0.4$\\
  \hline
 $B_{u}^{-}$ ${\to}$ ${\pi}^{0}K^{{\ast}-}$ & $6.9\pm2.3$ & $4.2\pm0.2$ & $4.4\pm0.4$ & $4.4\pm0.4$ & $4.3\pm0.3$ & $4.3\pm0.3$\\
 ${\overline{B}}_{d}^{0}$ ${\to}$ ${\pi}^{0}{\overline{K}}^{{\ast}0}$ & $2.4\pm0.7$ & $3.4\pm0.3$ & $3.5\pm0.2$ & $3.5\pm0.2$ & $3.1\pm0.3$ & $2.9\pm0.2$\\
  \hline
 $B_{u}^{-}$ ${\to}$ ${\rho}^{-}{\overline{K}}^{0}$ & $8.0^{+1.5}_{-1.4 }$ & --- & $8.6\pm0.7$ & $8.6\pm0.7$ & $7.4\pm0.4$ & $7.1\pm0.4$\\
 $B_{u}^{-}$ ${\to}$ ${\rho}^{0}K^{-}$ & $3.81^{+0.48}_{-0.46 }$ & $3.4\pm0.2$ & --- & --- & $3.4\pm0.2$ & $3.4\pm0.2$\\
 ${\overline{B}}_{d}^{0}$ ${\to}$ ${\rho}^{+}K^{-}$ & $8.6^{+0.9}_{-1.1 }$ & $9.7\pm0.5$ & --- & --- & $9.7\pm0.5$ & $9.8\pm0.5$\\
 ${\overline{B}}_{d}^{0}$ ${\to}$ ${\rho}^{0}{\overline{K}}^{0}$ & $5.4^{+0.9}_{-1.0 }$ & --- & $6.5\pm0.4$ & $6.5\pm0.4$ & $5.5\pm0.3$ & $5.4\pm0.4$\\
 \hline \hline
 \end{tabular}
 \end{center}
 \end{table}
%%%%%%%%%%%%%%%%%%%%%%%%%%%%%%%%%%%%%%%%%%%%

%%%%%%%%%%%%%%%%%%%%%%%%%%%%%%%%%%%%%%%%
\begin{table}[ht]
 \begin{center}
 \caption{The direct CP asymmetries ( in unit of $10^{-2}$) of $B$
 ${\to}$ $\pi K$, $\pi K^{\ast}$ and $\rho K$ decays. Other captions are the same as Table \ref{tabNPbr}}
 \label{tabNPcp}
 %\vspace{0.5cm}
 \doublerulesep 0.7pt \tabcolsep 0.07in
 \begin{tabular}{lccccccccccc} \hline \hline
 \multicolumn{1}{c}{Decay Mode}&\multicolumn{1}{c}{Experiment} &\multicolumn{5}{c}{NP}\\
 & \multicolumn{1}{c}{data}& \multicolumn{1}{c}{Case I} & \multicolumn{1}{c}{Case II}& \multicolumn{1}{c}{Case III}& \multicolumn{1}{c}{Case IV}& \multicolumn{1}{c}{Case V}\\ \hline
 $B_{u}^{-}$ ${\to}$ ${\pi}^{-}{\overline{K}}^{0}$ & $0.9\pm2.5$ & --- & $1.7\pm2.9$ & $2.0\pm0.2$ & $3.9\pm1.0$ & $3.2\pm1.3$\\
 $B_{u}^{-}$ ${\to}$ ${\pi}^{0}K^{-}$ & $5.0\pm2.5$ & $8.8\pm6.4$ & $1.1\pm0.9$ & $1.2\pm0.9$ & $2.8\pm5.5$ & $1.8\pm1.3$\\
 ${\overline{B}}_{d}^{0}$ ${\to}$ ${\pi}^{+}K^{-}$ & $-9.7\pm1.2$ & $-5.7\pm4.4$ & --- & ---& $-10.0\pm0.8$ & $-9.2\pm1.3$\\
 ${\overline{B}}_{d}^{0}$ ${\to}$ ${\pi}^{0}{\overline{K}}^{0}$ & $-14\pm11$ & $-18.6\pm7.5$ & $-12.8\pm3.9$ & $-12.6\pm1.6$ & $-10.2\pm7.0$ & $-8.2\pm2.8$\\
  \hline
 $B_{u}^{-}$ ${\to}$ ${\pi}^{0}K^{{\ast}-}$ & $4\pm29$ & $4.2\pm19.3$ & $-8.1\pm3.3$ & $-8.0\pm3.3$ & $-4.9\pm19.7$ & $-13.2\pm4.6$\\
 ${\overline{B}}_{d}^{0}$ ${\to}$ ${\pi}^{0}{\overline{K}}^{{\ast}0}$ & $-9^{+32}_{-23}$ & $-61.7\pm22.0$ & $-49.9\pm3.4$ & $-49.8\pm3.8$ & $-52.8\pm24.2$ & $-47.0\pm6.5$\\
  \hline
 $B_{u}^{-}$ ${\to}$ ${\rho}^{-}{\overline{K}}^{0}$ & $-12\pm17$ & --- & $-5.9\pm10.9$ & $-6.5\pm0.8$ & $-15.1\pm4.2$ & $-13.1\pm5.9$\\
 $B_{u}^{-}$ ${\to}$ ${\rho}^{0}K^{-}$ & $37\pm11 $ & $32.8\pm16.5$ & --- & --- & $48.3\pm3.5$ & $43.9\pm5.2$\\
 ${\overline{B}}_{d}^{0}$ ${\to}$ ${\rho}^{+}K^{-}$ & $15\pm13$ & $19.2\pm12.9$ & --- & --- & $31.9\pm2.7$ & $28.0\pm4.1$\\
 ${\overline{B}}_{d}^{0}$ ${\to}$ ${\rho}^{0}{\overline{K}}^{0}$ & $-2\pm29$ & --- & $-8.1\pm8.1$ & $-8.5\pm0.9$ & $-14.9\pm3.0$ & $-13.5\pm4.4$\\
 \hline \hline
 \end{tabular}
 \end{center}
 \end{table}
%%%%%%%%%%%%%%%%%%%%%%%%%%%%%%%%%%%%%%%%%%%%

%%%%%%%%%%%%%%%%%%%%%%%%%%%%%%%%%%%%%%%%
\begin{table}[ht]
 \begin{center}
 \caption{ The numerical results for the parameters $C_{S1}^{u}$, $C_{S1}^{u}$, $\delta_{S}^{u}$, $C_{S1}^{d}$, $C_{S8}^{d}$ and $\delta_{S}^{d}$ in different NP Cases. The dashes mean the corresponding operators are neglected in the  Case.}
 \label{NPSpace}
 \vspace{0.5cm}
 \doublerulesep 0.7pt \tabcolsep 0.07in
 \begin{tabular}{lccccccccccc} \hline \hline
 NP para.&Case I&Case II&Case III&Case IV&Case V&\\ \hline
 $C_{S1}^{u}$$(\times10^{-3})$& $-41.6\pm13.4$ & --- & --- & $25.8\pm8.4$ & $-6.7\pm10.5$\\
 $C_{S8}^{u}$$(\times10^{-3}$)& $38.7\pm18.2$ & --- & --- & --- & $16.0\pm7.1$ \\
 $\delta_{S}^{u}$ & $99.5^{\circ}\pm6.1^{\circ}$ & --- & --- & $107.0^{\circ}\pm11.5^{\circ}$ & $73.0^{\circ}\pm23.8^{\circ}$ \\ \hline
  $C_{S1}^{d}$$(\times10^{-3})$& --- & $23.0\pm5.1$ & $22.8\pm2.3$ & $50.3\pm12.8$ & $17.5\pm10.1$\\
 $C_{S8}^{d}$$(\times10^{-3})$& --- & $-0.8\pm13.7$ & --- & --- & $10.5\pm9.4$\\
 $\delta_{S}^{d}$ & --- & $100.0^{\circ}\pm8.7^{\circ}$ & $99.3^{\circ}\pm9.2^{\circ}$ & $106.6^{\circ}\pm7.3^{\circ}$ & $114.7^{\circ}\pm18.6^{\circ}$\\
 \hline \hline
 \end{tabular}
 \end{center}
 \end{table}
%%%%%%%%%%%%%%%%%%%%%%%%%%%%%%%%%%%%%%%%%%%%
{\bf  Case I: $b\to s u{\bar u}$ operators $O^u_{S1}$ and $O^u_{S8}$}

We just take into account the contributions of $O_{S1}^{u}$ and
$O_{S8}^{u}$ in Eq.~(\ref{eq:np}), \textit{i.e.}
$C_{S1}^{d}=C_{S8}^{d}=0$. In this case, we take the branching
ratios of the seven relevant decays $B_{u}^{-}{\to}{\pi}^{0}K^{-}$,
${\pi}^{0}K^{\ast -}$, ${\rho}^{0}K^{-}$ and
${\overline{B}}_{d}^{0}{\to}{\pi}^{+}K^{-}$, ${\pi}^{0}\bar{K}^{0}$,
${\pi}^{0}\bar{K}^{\ast 0}$, ${\rho}^{+}K^{-}$ as constraints and
leave the direct CP asymmetries as our predictions. The allowed
regions of the NP parameters $C_{S1}^{u}$, $C_{S8}^{u}$ and
$\delta_{S}^{u}$ are shown in Fig.~\ref{CaseI}.  From which, we find the spaces of $C_{S1}^{u}$ and
$\delta_{S}^{u}$  consist  of two parts (dark and gray). However,
with the gray part, we get $A_{CP}(B_{u}^{-}$ ${\to}$
${\pi}^{0}K^{-})=-0.154\pm0.038$ which conflicts with experimental
data $0.050\pm0.025$. So, the gray region should be excluded. With
the dark part of parameter spaces, our prediction $A_{CP}(B_{u}^{-}$
${\to}$ ${\pi}^{0}K^{-})=0.088\pm0.064$ is consistent with
experimental data. Furthermore, the branching ratios and direct CP
asymmetries of the other decay modes, listed in the third column of
Table~\ref{tabNPbr} and \ref{tabNPcp}, agree with experimental data
within error bars. The constrained  parameter space  $C_{S1}^{u}$,
$C_{S8}^{u}$ and $\delta_{S}^{u}$ are listed in the second column of
Table~\ref{NPSpace}. We note that
$C_{S1}^{u}\approx-C_{S8}^{u}\approx-0.04$ with
$\delta_{S}^{u}\approx100^{\circ}$, it means the strength of
color-singlet and color-octet operators are similar, however, such a situation  may be hard
to be generated with a realistic  available  NP model.
%%%%%%%%%%%%%%%%%%%%%%%%%%%%%%%%%%%%%%%%
%%%%%%%%%%%%%%%%%%%%%%%%%%%%%%%%%%%%%%%%%%%
\\
\\
\textbf{Case II: $b\to s d{\bar d}$ operators $O^d_{S1}$ and $O^d_{S8}$ }

In a large category of  NP scenarios with scalar interactions, for
example, two-Higgs doublets model II,  down type fermion  Yukawa
couplings are enhanced.  So, in this case, we evaluate the effects
of $O_{S1}^{d}$ and $O_{S8}^{d}$ and neglect  $O_{S1}^{u}$ and
$O_{S8}^{u}$.

As shown by Eqs.~(\ref{amp1_NP})--(\ref{amp12_NP}), $O_{S1(8)}^{d}$
contributes to the decays $B{\to}{\pi}^{-}\bar{K}^{0}$,
${\pi}^{0}K^{-}$, ${\pi}^{0}K^{\ast -}$, ${\rho}^{-}\bar{K}^{0}$,
${\pi}^{0}\bar{K}^{0}$, ${\pi}^{0}\bar{K}^{\ast 0}$, and
${\rho}^{0}\bar{K}^{0}$. From Table \ref{tab_br}, one can find
that the SM predictions for their branching ratios are consistent
with the experimental data. So, in this Case, NP weak phase
$\delta_{S}^{d}$ would be arbitrary for  very small strengths  of
$C_{S1}^{d}$ and $C_{S8}^{d}$,  we  thus have to take into account
both branching ratios and direct CP violations  as constraints. The
allowed region of $C_{S1}^{d}$, $C_{S8}^{d}$ and $\delta_{S}^{d}$
are shown in Fig.~\ref{CaseII}. The fitted results are shown in
the fourth column of Table~\ref{tabNPbr}, \ref{tabNPcp} and the third
column of Table~\ref{NPSpace}. Interestingly, we note that
$C_{S1}^{d}=0.023\pm0.005$, $C_{S8}^{d}=-0.001\pm0.013$ (consistent
with zero) with $\delta_{S}^{d}\approx100^{\circ}$. It indicates
that color-singlet operator $O_{S1}^{d}$ dominates the NP $b\to
sd\bar{d}$ contributions. Actually, with $O_{S1}^{d}$ only, we could
find a solution to the ``$\pi K$ puzzle'' which will be discussed in
next Case.

Compared with Case I, it is found that
$|C_{S1}^{d}|<|C_{S1}^{u}|\approx|C_{S8}^{u}|$. However, we can't
conclude that $O_{S1(8)}^{u}$ dominates the NP contribution  until
we consider the two operators simultaneously, which will be
discussed in coming Case IV and Case V.

%%%%%%%%%%%%%%%%%%%%%%%%%%%%%%%%%%%%%%%%
%%%%%%%%%%%%%%%%%%%%%%%%%%%%%%%%%%%%%%%%%%%

\textbf{Case III: $b\to s d{\bar d}$ operator $O^d_{S1}$ solely }

As the former Case, both branching ratio and direct CP violation are
taken as constraints. With $O_{S1}^{d}$ solely, we find a solution
to the ``$\pi K$ puzzle'' with the $C_{S1}^{d}$ and $\delta_{S}^{d}$
allowed region shown in Fig.~\ref{CaseIII}. The numerical results
are listed in fifth column of Table~\ref{tabNPbr}, \ref{tabNPcp} and
fourth column of Table~\ref{NPSpace}, respectively.  $C_{S1}^{d}$
and $\delta_{S}^{d}$ are found similar to the ones of Case II. It
confirms our findings in Case II that $O_{S1}^{d}$ dominates the NP
contributions and the contribution of $O_{S8}^{d}$ is negligible. As
known, it is easy to generate the situation in  many NP scenarios.
However, both the strength $C_{S1}^{d}\approx 0.022$  and the new
weak phase $\delta^{d}\approx99^{\circ}$ normalized to
$\frac{G_{F}}{\sqrt{2}} |V_{tb}V_{ts}^{\ast}|$ may be  toughly large
for realistic NP models without violating other precise electro-weak
measurements.

%%%%%%%%%%%%%%%%%%%%%%%%%%%%%%%%%%%%%%%%
%%%%%%%%%%%%%%%%%%%%%%%%%%%%%%%%%%%%%%%%%%%

\textbf{Case IV:   only color-singlet operators  $O^u_{S1}$ and $O^d_{S1}$  }

In order to compare the relative strength of two color singlet
operators $O_{S1}^{d}$ and $O_{S1}^{u}$, we take them into account
at the same time and neglect the other two color-octet ones. Taking
the branching ratios of the relevant  decays as constraints, we find
the allowed  regions for the NP parameters $C_{S1}^{u}$,
$\delta_{S}^{u}$, $C_{S1}^{d}$ and $\delta_{S}^{d}$, which  are
shown in Fig.~\ref{CaseIV}.  All  our predictions for the direct CP
violations, listed in in sixth column of Table~\ref{tabNPcp}, agree
with experimental data. Especially, we note our
predictions  $A_{CP}(B^{-}{\to}{\pi}^{0}K^{-})=0.028\pm0.055$ and
$\Delta A=0.128\pm 0.056$ agree with experimental data very well.

The fifth column of Table~\ref{NPSpace} is the parameter space
obtained for the present Case. We find  that strength of
$C_{S1}^{d}$ in Case IV is larger than the ones in Case II and Case
III, because the terms of $C_{S1}^{d}$ and $C_{S1}^{u}$ always have
opposite sign in
Eqs.~(\ref{amp2_NP}),~(\ref{amp4_NP}),~(\ref{amp6_NP})
and~(\ref{amp8_NP}), but only one of them exists in the other decay
modes. It is found that $C_{S1}^{d}\approx 2\times
C_{S1}^{u}\approx0.05$ with
$\delta_{S}^{d}\approx\delta_{S}^{u}\approx107^{\circ}$, which shows
$O_{S1}^{d}$ dominance.

%%%%%%%%%%%%%%%%%%%%%%%%%%%%%%%%%%%%%%%%
%%%%%%%%%%%%%%%%%%%%%%%%%%%%%%%%%%%%%%%%%%%

%%%%%%%%%%%%%%%%%%%%%%%%%%%%%%%%%%%%%%%%
%%%%%%%%%%%%%%%%%%%%%%%%%%%%%%%%%%%%%%%%%%%

\textbf{ Case V: all the operators $O^u_{S1}$, $O^u_{S8}$,
$O^d_{S1}$ and $O^d_{S8 }$ }

At last, we fit the measured branching ratios and the direct CP
violations of all the relevant ten decay models with  the four
operators in Eq.~(\ref{eq:sp}). Generally the ten CP averaged
branching ratios are measured with high significants, however, only
$ A_{CP}(B^{0} \to \pi^{\pm}K^{\mp})$ is well established at
$8\sigma$ level and $A_{CP}(B_{u}^{-}{\to}{\pi}^{0}K^{-})$ by itself
is at $2\sigma$ level only.

From the fit, the allowed regions for the six NP parameters
$C_{S1}^{u}$, $C_{S8}^{u}$ $\delta_{S}^{u}$, $C_{S1}^{d}$,
$C_{S8}^{d}$ and $\delta_{S}^{d}$ shown in Fig.~\ref{CaseV}. The
fitted branching ratios and CP violations are listed in the seventh
column of Table~\ref{tabNPbr} and \ref{tabNPcp}, and the  fitted
values of the NP parameters are presented in the last column of
Table~\ref{NPSpace}, respectively. Since the experimental data are
allowed varying randomly within their $2\sigma$ error-bars,  the
uncertainties of our fitting results are turned to be quite large.

We find $C_{S1}^{u}=(-6.7\pm10.5)\times10^{-3}$ and
$C_{S8}^{u}=(16.0\pm7.1)\times10^{-3}$ with
$\delta^{u}_{S}=73.0^{\circ}\pm23.8^{\circ}$, which shift our
predication $A_{CP}( \bar{B}^{0}\to \pi^{\pm}K^{\mp})\approx-0.124$
in the SM more closer to the experimental data $-0.097$. However, it
does not indicate that the $b\to su\bar{u}$ operators are important for
resolving CP violation difference $\Delta A$, since the sum of their
contributions  to $B^{-}\to \pi^{0}K^{-}$ is quite small due to
cancellation among them. For the $b\to s \bar{d}d$ operators, we get
$C_{S1}^{d}=(17.5\pm10.1)\times10^{-3}$ and
$C_{S8}^{d}=(10.5\pm9.4)\times10^{-3}$ with
$\delta^{d}_{S}=114.7^{\circ}\pm18.6^{\circ}$. The results  are
consistent  with  these of Case II and Case III as shown in
Table~\ref{NPSpace}, however, due to interferences with $b\to
su\bar{u}$ contributions, the uncertainties are much larger than the
two former Cases where $b\to su\bar{u}$ operators  are dropped.
Moreover, as shown by Eq.~(\ref{amp2_NP}), $C_{S8}^{d}$ is
suppressed by $1/N_{c}$ in the amplitude of $B^{-}\to\pi^{0}K^{-}$,
thus, the dominate status of $O_{S1}^{d}$ for resolving $\Delta A$
is remained.

\subsection{The mixing-induced CP asymmetries in $B\to\pi^{0}K_{S}$ and $B\to\rho^{0}K_{S}$ }
%%%%%%%%%%%%%%%%%%%%%%%%%%%%%%%%%%%%%%%%
\begin{table}[ht]
 \begin{center}
 \caption{The mixing-induced CP asymmetries ( in unit of $10^{-2}$) of $\bar{B}^0\to\pi^{0}K_{S}\,,\rho^{0}K_{S}$ decays. Other captions are the same as Table \ref{tabNPbr}}
 \label{tabNPmixcp}
 \vspace{0.5cm}
 \doublerulesep 0.7pt \tabcolsep 0.07in
 \begin{tabular}{lccccccccccc} \hline \hline
 \multicolumn{1}{c}{Decay Mode}&\multicolumn{1}{c}{Experiment} &\multicolumn{1}{c}{SM} &\multicolumn{5}{c}{NP}\\
 & \multicolumn{1}{c}{data}& &\multicolumn{1}{c}{Case I} & \multicolumn{1}{c}{Case II}& \multicolumn{1}{c}{Case III}& \multicolumn{1}{c}{Case IV}& \multicolumn{1}{c}{Case V}\\ \hline
 ${\overline{B}}_{d}^{0}$ ${\to}$ ${\pi}^{0}K_S$ & $38\pm19$ & $77\pm4$ & $45\pm11$ & $56\pm5$ & $57\pm3$ & $59\pm9$& $62\pm8$\\
 ${\overline{B}}_{d}^{0}$ ${\to}$ ${\rho}^{0}K_S$ & $61^{+25}_{-27}$ & $66\pm3$ &---& $61\pm6$ & $61\pm3$ & $56\pm3$& $57\pm4$\\
 \hline \hline
 \end{tabular}
 \end{center}
 \end{table}
%%%%%%%%%%%%%%%%%%%%%%%%%%%%%%%%%%%%%%%%%%%%

So far we have discussed the direct CP asymmetries in the these decays with five NP scenarios.  However, it is naturally to 
question if we can account for the mixing-induced CP asymmetries in $\bar{B}^0\to\pi^{0}K_S$ and $\rho^{0}K_{S}$
 decays with these constrained parameter spaces obtained in the former subsection.   
As known, the mixing-induced asymmetries  are  more suitable  for probing new physics effects entered via $b\to s q{\bar q}$
parton processes than the direct ones, since the former ones could be predicted more accurately in QCDF.  Detail discussions
 for the interesting feature  could be found in Ref. \cite{benekeCP}.  Recently,  the measured relative small  mixing-induced asymmetry ( with large error-bar ) in $\bar{B}^0\to\pi^0 K_S$  has attracted much attention in the literature
\cite{benekeCP,Nir,Fleischer,mannel,rosner}. 

The time-dependent CP asymmetries in $\bar{B}^0\to\pi^{0}K_{S}$ and $\rho^{0}K_S$
 decays could be written as 
 \begin{equation}
 {\cal A}_{f}(t)=S_{f} \sin(\Delta m_{d}t) - C_{f} \cos (\Delta m_{d}t),
 \end{equation}
 where $-C_{f}\equiv {\cal A}_{CP}$ is the direct CP violation already discussed in former subsection.  $S_{f}={\cal A}^{mix}_{CP}$ is  the mixing-induced  asymmetry   
 \begin{equation}
 A^{mix}_{CP}(\bar{B}^0\to f)=\frac{2{\rm Im}\lambda_f}{1+|\lambda_f|^2}\qquad
 (f=\pi^0 K_S\,,\rho^0 K_S, ~~~\eta_{f}=-1 )
\end{equation}
%%%%%%%%%%%%%%%%%%%%%%%%%%%%%%%%%%%%%%%%%%
where $\lambda_f=-e^{-2i\beta}\bar{A}^{00}/A^{00}$ and $\sin(2\beta)=\sin(2\beta)_{\Psi K_{S}}=0.68\pm0.03$\cite{HFAG},  since the NP operators are irrelevant to $B^{0}-\bar{B}^{0}$ mixing amplitude. 
 
 Using the constrained parameters of the NP operators in Table \ref{NPSpace} and taking $m_g=0.5\,{\rm GeV}$, our numerical results are listed in Table~\ref{tabNPmixcp} for the SM and the five Cases of NP operators. The experimental data column is the  averages by HFAG~\cite{HFAG}. In the SM, up to doubly Cabibbo suppressed amplitudes, one can expect 
 \begin{equation} 
 {\cal A}_{CP}\approx0, ~~~~{\cal A}^{mix}_{CP}=S\approx \sin(2\beta)_{\Psi K_{S}}=0.68\pm0.03
 \end{equation}  
for the two decay modes.  We  get $A^{mix}_{CP}(\pi^{0}K_{S})=0.77\pm0.04$ and  $A^{mix}_{CP}(\rho^{0}K_{S})=0.66 
 \pm0.03$. It is noted that the former is slight larger than $\sin(2\beta)_{\Psi K_{S}}$ which is due to corrections of the 
  suppressed amplitudes proportional to $V_{ub}V_{us}^{*}$ as discussed in Ref.\cite{benekeCP} \footnote{ If the old data
   $\sin(2\beta)_{\Psi K_{S}}=0.725\pm0.037$ used, we get $\Delta S_{\pi^{0}K_{s}}= S_{\pi^{0}K_{s}}-\sin(2\beta)_{\Psi K_{S}}
   =0.05\pm0.08$ and  $\Delta S_{\rho^{0}K_{s}}=-0.05\pm0.07$, which agree well with the results in the paper. Considering our different treatments of the end-piont divergences, the agreement numerically confirms the observation  that the mixing induced CP violations are insensitive to strong phases in the decay amplitudes.}.   As shown in Table.\ref{tabNPmixcp}, the NP pseudoscalar operators  decrease $S_{\pi^{0}K_{S}}$ and $S_{\rho^{0}K_{S}}$ (weaker than former),  which seems to be
 favored by the experimental data.

\begin{figure}[ht]
\begin{center}
\epsfxsize=15cm \centerline{\epsffile{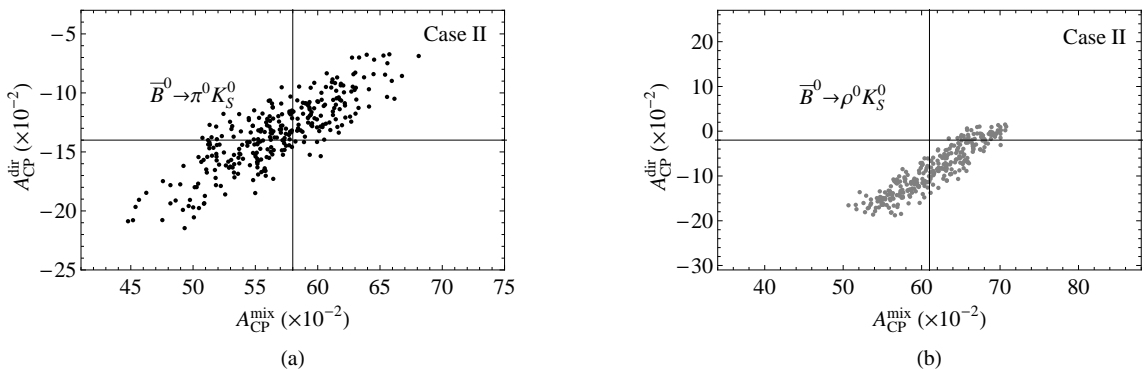}}
\centerline{\parbox{16cm}{\caption{\label{CPfig}\small Correlation
between direct and mixing-induced CP asymmetries for
$\bar{B}^0\to\pi^{0}K_{S}$ (a) and $\bar{B}^0\to\rho^{0}K_S$ (b) in Case II
with $m_g=0.5\,{\rm GeV}$. The lines are the central value of
experimental data presented at ICHEP 08. Our plot ranges corresponding the experimental error-bars.}}}
\end{center}
\end{figure}

  We note that HFAG has not included the following data yet   
%%%%%%%%%%%%%%%%%%%%%%%%%%%%%%%%%%%%%%%%
\begin{eqnarray}
&&A^{mix}_{CP}(\bar{B}^0\to\pi^0 K_S)=0.55\pm0.20\pm0.03\qquad {\rm BABAR}~\cite{mixcpbabar}\,, \\
&&A^{mix}_{CP}(\bar{B}^0\to\pi^0 K_S)=0.67\pm0.31\pm0.08\qquad
{\rm Belle}~\cite{mixcpbelle}\,,
\end{eqnarray}
%%%%%%%%%%%%%%%%%%%%%%%%%%%%%%%%%%%%%%%%%%
which are reported very recently at ICHEP08. The average reads $A^{mix}_{CP}(\bar{B}^0\to\pi^0
K_{S})=0.58\pm0.17$.  Again from  Table \ref{tabNPmixcp}, we find the outputs of all the five Cases with their fitted parameter spaces  are in good agreements with the  new experimental results since the error-bar are still large.   Taking Case II as example, i.e., assuming NP from $b\to s d{\bar d}$, we present the correlations of the direct and the mixing-induced
CP asymmetries for $\bar{B}^0\to\pi^0K_{S}$ and
$\bar{B}^0\to\rho^{0}K_{S}$ decays in Fig.~\ref{CPfig}, where the constrained parameters listed in  Table \ref{NPSpace} are 
used.  Although all points fall in the present experimental error-bars,  Fig.~\ref{CPfig} shows interesting correlations between 
$A^{dir}_{CP}$ and $A^{mix}_{CP}$(S).  If the experimental $S_{\pi^{0}K_{S}}$   shrank to be much lower than 
$\sin(2\beta)_{\Psi K_{S}}$, the NP  Case II would give large negative  direct  CP asymmetry.  Similar implication also applies to  $\rho^{0}K_{S}$ final states.       
%%%%%%%%%%%%%%%%%%%%%%%%%%%%%%%%%%%%%%%%
%%%%%%%%%%%%%%%%%%%%%%%%%%%%%%%%%%%%%%%%%%%

In summary, assuming NP effects entering $B\to \pi K, \pi K^{*}$ and
$\rho K$ decays via $\bar{s} (S+P) b \otimes \bar{q} (S+P) q$
operators, we have performed fittings for the observables in these
decays with a model-independent approach. It's found that all the
experimental data, especially the direct CP violation difference
$\Delta A$, could be accommodated by new $b\to su\bar{u}$ or $b\to
sd\bar{d}$ contributions, of course by their combination. Assuming
the dominance of  new $b\to su\bar{u}$ operators (Case I), we find
the color-octet operator has the similar strength as the
color-singlet one, which is  rather exotic  for  electro-weak NP
models. However, taking the new $b\to sd\bar{d}$ operators dominant
(Cases II and III), we have shown that color-singlet operator
$\bar{s} (S+P) b \otimes \bar{d} (S+P) d$ solely can provide a
resolution to the derivations with a strength about half of $b\to
su\bar{u}$ operators. We also have  performed fits (Cases IV and V)
with both $b\to su\bar{u}$ and $b\to sd\bar{d}$ contributions to
infer the their relative size in these decays. It is found that the
strength of $b\to sd\bar{d}$ is stronger than that of $b\to
su\bar{u}$. In all cases, to account for the experimental deviations
from the SM predictions for direct CP violations, especially for
$A_{CP}(B^{-}{\to}{\pi}^{0}K^{-})$, new electro-weak phase about
$100^{\circ}$ relative to the SM  $b\to sq\bar{q}$ penguin amplitude
is always required. With the fitted parameters, we present results for the mixing 
induced CP asymmetries in $\bar{B}^0\to\pi^{0}K_S$ and $\rho^{0}K_S$ decays.
It is found the NP effects generally reduce $S_{\pi^{0}K_S}$ and $S_{\rho^{0}K_{S}}$. However,
 due to the large error-bars, the present experimental data do not further reduce the parameter 
 spaces of the NP operators.    
 
\section{Conclusions}
At present, the successful running of the  B factories with their
detectors BABAR (SLAC) and BELLE (KEK) have already taken about
$10^{9}$ data together at $\Upsilon(4S)$ resonance, and have
produced plenty of exciting results. Tensions between the
experimental data and the SM predictions based on different
approaches for strong dynamics  are accumulated, which may be due to
our limited understanding of the strong dynamics, but equally
possible due to NP effects. Motivated  by the  recent observed
$\Delta A$ of the difference in direct CP violation between  ${\cal
A}_{CP}(B^{\mp}\to\pi^0 K^{\mp})$ and $ A_{CP}( B^{0}\to
K^{\pm}\pi^{\mp})$ and theoretical issues of end-point divergences,
strong phases and  annihilation contributions in charmless hadronic
B decays, we have revisited the $B\to\pi K, \pi K^{\ast}$ and $\rho
K$ decays with an infrared finite form of the gluon propagator
supplemented to the QCDF approach. In this way, we can get large
strong phases from the annihilation contributions, while the hard
spectator-scattering amplitudes are real. From our numerical
analyses, we find that the contributions of the annihilation and the
hard-spectator topologies are sensitive to the value of the
effective gluon mass $m_g$. With $m_g=500\pm50~{\rm MeV}$, our
predictions in the SM agree with the current experimental data well,
except $A_{CP}(B^{\pm}\to K^{\pm}\pi^{0})$. Actually with $m_g$
varying from $300~{\rm MeV}$ to $700~{\rm MeV}$, we always get
${\cal A}_{CP}(B^{\mp}\to\pi^0 K^{\mp}) \approx A_{CP}( B^{0}\to
K^{\pm}\pi^{\mp})$ as shown in Table~\ref{tab_cp}, which also agree
with the results in the literature. We conclude that NP effects is
required, at least can not be excluded, to resolve the discrepancies
between the observed $\Delta A$ and the SM expectations.

With four effective NP $b\to su\bar{u}$ and $b\to sd\bar{d}$
operators, we have performed a model-independent approach to the
discrepancies. Our main conclusions are summarized as:

\begin{itemize}
\item Assuming dominance of $b\to su\bar{u}$ operators,
the fit gives a quite small center value for $A_{CP}( B^{0}\to
K^{\pm}\pi^{\mp})$ although consistent with the data within its
large error-bar. Moreover, the strength of color-octet operator
$O_{S8}^{u}$ is comparable with color-singlet $O_{S1}^{u}$ which may
be  rather exotic for most NP models.
\item With  the $b\to sd\bar{d}$ operator $O_{S1}^{d}$ solely,
the observables in $B\to\pi K, \pi K^{\ast}$ and $\rho K$ decays
could be well accommodated, since $\bar{B}^{0}\to K^{-}\pi^{+}$ is
irrelevant to the $b\to sd\bar{d}$ operator and it's branching ratio
and CP violation agree with the SM prediction very well.
\item  Assuming dominance of color-singlet operators $O_{S1}^{u}$ and  $O_{S1}^{d}$,
it is found that the two operators have the similar weak
phase with $C_{S1}^{u} \approx \frac{1}{2} C_{S1}^{d}$.
\item For all Cases, to account for the experimental
deviations from the SM predictions for direct CP violations,
especially for  $A_{CP}(B^{-}{\to}{\pi}^{0}K^{-})$, new electro-weak
phase about $100^{\circ}$ relative to the SM $b\to sq\bar{q}$
penguin amplitude is always required.
\item With the fitted parameter spaces, the NP operators decrease the mixing-induced CP violations in 
 $B^{0} \to \pi^{0}K_S$ and $\rho^{0}K_{S}$ decays, especially that of $\pi^{0}K_S$ final states.     
\end{itemize}

It is reminded that both direct and mixing-induced  CP violations have not been well
established in most of charmless nonleptonic B decays. Although  the difference in direct CP asymmetries  between  
${\cal A}_{CP}(B^{\mp}\to\pi^0 K^{\mp})$ and $ A_{CP}( B^{0}\to K^{\pm}\pi^{\mp})$ shows some hints of new physics activities,  we still need refined measurements of the mixing-induced CP asymmetries in the related decays $B^{0}\to\pi^{0}K_{S}$ and $\rho^{0}K_{S}$ to confirm or refute the NP hints, since the former  strongly depends on strong phases in the decay
amplitudes while the later not so much and can be predicted more precisely.  In the coming
years, the precision of experimental  measurement of the observables in
these decays will be improved much with LHCb at CERN, which will
shrink the parameter space and reveal the relative importance of the
five Cases studied in this paper. Then, the favored Case  will
deserve detail studies with particular  NP models.

%%%%%%%%%%%%%%%%%%%%%%%%%%%%%%%%%%%%%%%%%%%%
\section*{Acknowledgments}
The work is supported by National Science Foundation under contract
Nos.10675039 and 10735080. X.Q. Li acknowledges support from the
Alexander-von-Humboldt Stiftung.

%%%%%%%%%%%%%%%%%%%%%%%%%%%%%%%%%%%%%%%%%%%%
\begin{appendix}

\section*{Appendix~A: Decay amplitudes in the SM with QCDF}
The amplitudes for $B\to\pi K$, $\pi K^{\ast}$ and $\rho K$ are
recapitulated from Ref.~\cite{Beneke3}
\begin{eqnarray}
{\cal A}_{B^-\to\pi^- \bar{K}}^{\rm SM}
   &=& \sum_{p=u,c}V_{pb}V_{ps}^{\ast} A_{\pi \bar{K}} \Big[
    \delta_{pu}\,\beta_2 + \alpha_4^p - \half\alpha_{4,{\rm
    EW}}^p +\beta_3^p+\beta_{3,{\rm
    EW}}^p\Big],
\label{amp1_SM}
\end{eqnarray}
\begin{eqnarray}
\sqrt2\, {\cal A}_{B^-\to\pi^0 K^-}^{\rm SM}
   &=& \sum_{p=u,c}V_{pb}V_{ps}^{\ast} \biggl\{ A_{\pi^0 K^-} \Big[
    \delta_{pu}\,(\alpha_1+\beta_2) + \alpha_4^p + \alpha_{4,{\rm
    EW}}^p +\beta_3^p+\beta_{3,{\rm EW}}^p\Big]\nonumber\\
   &&+  A_{ K^- \pi^0}\Big[\delta_{pu}\,\alpha_2+\frac{3}{2}\alpha_{3,{\rm EW}}^p\Big]\biggl\},
\label{amp2_SM}
\end{eqnarray}
\begin{eqnarray}
{\cal A}_{\bar{B}^0\to\pi^+ K^-}^{\rm SM}
   &=& \sum_{p=u,c}V_{pb}V_{ps}^{\ast} A_{\pi^+ K^-} \Big[
    \delta_{pu}\,\alpha_1 + \alpha_4^p + \alpha_{4,{\rm
    EW}}^p +\beta_3^p-\half\beta_{3,{\rm EW}}^p\Big],
\label{amp3_SM}
\end{eqnarray}
\begin{eqnarray}
\sqrt2\, {\cal A}_{\bar{B}^0\to\pi^0 \bar{K}^0}^{\rm SM}
   &=& \sum_{p=u,c}V_{pb}V_{ps}^{\ast} \biggl\{ A_{\pi^0 \bar{K}^0} \Big[
    -\alpha_4^p + \half\alpha_{4,{\rm
    EW}}^p -\beta_3^p+\half\beta_{3,{\rm EW}}^p\Big]\nonumber\\
   &&+  A_{ \bar{K}^0 \pi^0}\Big[\delta_{pu}\,\alpha_2+\frac{3}{2}\alpha_{3,{\rm
   EW}}^p\Big]\biggl\},
\label{amp4_SM}
\end{eqnarray}
where the explicit expressions for the coefficients
$\alpha_i^p\equiv\alpha_i^p(M_1M_2)$ and
$\beta_i^p\equiv\beta_i^p(M_1M_2)$ can also be found in
Ref.~\cite{Beneke3}. Note that expressions of the hard spectator
terms $H_i$ appearing in $\alpha_i^p$ and the weak annihilation
terms appearing in $\beta_i^p$ should be replaced with  our recalculated
ones. The amplitudes of $B\to\pi K^{\ast}$ and $B\to\rho K$ decays
could be obtained by setting $(\pi K)\to (\pi K^{\ast})$ and $(\pi K)\to (\rho K)$, respectively.

\section*{Appendix~B: Theoretical input parameters}

\subsection*{B1. Wilson coefficients and CKM matrix elements}
The Wilson coefficients $C_{i}(\mu)$ have been evaluated reliably to
next-to-leading logarithmic order~\cite{Buchalla:1996vs,Buras:2000}.
Their numerical results in the naive dimensional regularization
scheme at the scale $\mu=m_{b}$~($\mu_h=\sqrt{\Lambda_h m_b}$) are
given by
\begin{eqnarray}
&&C_{1}=1.074~(1.166), \quad C_{2}=-0.170~(-0.336), \quad
  C_{3}=0.013~(0.025), \nonumber \\
&&C_{4}=-0.033~(-0.057), \quad C_{5}=0.008~(0.011), \quad
  C_{6}=-0.038~(-0.076),\nonumber \\
&&C_{7}/\alpha_{e.m.}=-0.016~(-0.037), \quad
C_{8}/\alpha_{e.m.}=0.048~(0.095), \quad
  C_{9}/\alpha_{e.m.}=-1.204~(-1.321), \nonumber \\
&&C_{10}/\alpha_{e.m.}=0.204~(0.383), \quad
C_{7\gamma}=-0.297~(-0.360), \quad C_{8g}=-0.143~(-0.168).
\end{eqnarray}
The values at the scale $\mu_{h}$, with $m_{b}=4.80~{\rm GeV}$ and
$\Lambda_{h}=500~{\rm MeV}$, should be used in the calculation of
hard-spectator and weak annihilation contributions.

For the CKM matrix elements, we adopt the Wolfenstein
parameterization~\cite{Wolfenstein:1983yz} and choose the four
parameters $A$, $\lambda$, $\rho$, and $\eta$
as~\cite{Charles:2004jd}
\begin{equation}
A=0.807\pm 0.018, \quad \lambda=0.2265\pm 0.0008, \quad
\overline{\rho}=0.141^{+0.029}_{-0.017}, \quad
\overline{\eta}=0.343\pm0.016,
\end{equation}
with $\overline{\rho}=\rho\,(1-\frac{\lambda^2}{2})$ and
$\bar{\eta}=\eta\,(1-\frac{\lambda^2}{2})$.

\subsection*{B2. Quark masses and lifetimes}
As for the quark mass, there are two different classes appearing in
our calculation. One type is the pole quark mass appearing in the
evaluation of penguin loop corrections, and denoted by $m_q$. In
this paper, we take
\begin{equation}
 m_u=m_d=m_s=0, \quad m_c=1.64\pm0.09\,{\rm GeV}, \quad m_b=4.80\pm0.08\,{\rm GeV}.
\end{equation}
The other one is the current quark mass which appears in the factor
$r_\chi^M$ through the equation of motion for quarks. This type of
quark mass is scale dependent and denoted by $\overline{m}_q$. Here
we take~\cite{PDG06,HPQCD:2006}
\begin{eqnarray}
&&\overline{m}_s(\mu)/\overline{m}_q(\mu)=27.4\pm0.4~\cite{HPQCD:2006}\,,\quad
\overline{m}_{s}(2\,{\rm GeV}) =87\pm6\,{\rm MeV}~\cite{HPQCD:2006}\,,\nonumber\\
&&\overline{m}_{b}(\overline{m}_{b})=4.20\pm0.07{\rm
GeV}~\cite{PDG06}\,,
\end{eqnarray}
where $\overline{m}_q(\mu)=(\overline{m}_u+\overline{m}_d)(\mu)/2$,
and the difference between $u$ and $d$ quark is not distinguished.

As for the lifetimes of B mesons, we take~\cite{PDG06} $\tau_{B_{u}}
= 1.638\,{\rm ps}$ and $ \tau_{B_{d}}=1.530\,{\rm ps}$ as our
default input values.

\subsection*{B3. The decay constants and form factors}
In this paper, we take the decay constants
\begin{eqnarray}
 & &f_{B}=(216\pm22)~{\rm MeV}~\cite{Gray:2005ad}, \quad
    f_{B_s}=(259\pm32)~{\rm MeV}~\cite{Gray:2005ad}, \quad
    f_\pi=(130.7\pm0.4)~{\rm MeV}~\cite{PDG06},\nonumber\\
 & &f_{K}=(159.8\pm1.5)~{\rm MeV}~\cite{PDG06} \quad
    f_{K^{\ast}}=(217\pm5)~{\rm MeV}~\cite{BallZwicky}, \quad
    f_{\rho}=(209\pm2)~{\rm MeV}~\cite{PDG06}.
\end{eqnarray}
and the form factors~\cite{BallZwicky}
%%%%%%%%%%%%%%%%%%%%%%%%%%%%%%%%5
\begin{eqnarray}
 & &F^{B\to \pi}_{0}(0)=0.258\pm0.031, \quad
     F^{B\to {K}}_{0}(0)=0.331\pm0.041,\quad
     V^{B\to K^\ast}(0)=0.411\pm0.033,\nonumber\\
 & & A_0^{B\to K^\ast}(0)=0.374\pm0.034,\quad
     A_1^{B\to K^\ast}(0)=0.292\pm0.028,\quad
     V^{B\to \rho}(0)=0.323\pm0.030, \nonumber\\
  & &A_0^{B\to \rho}(0)=0.303\pm0.029,\quad
     A_1^{B\to \rho}(0)=0.242\pm0.023.
\end{eqnarray}
%%%%%%%%%%%%%%%%%%%%%%%%%%%%%%%%

\subsection*{B4. The LCDAs of mesons and light-cone projector operators.}
The light-cone projector operators of light pseudoscalar and vector
meson in momentum space read~\cite{Terentev,Beneke3}
%%%%%%%%%%%%%%%%%%%%%%%%%%%%%%%%%%%%%%%%%%%%
\begin{eqnarray}
M_{\alpha\beta}^{P}&=&\frac{if_P}{4}\Big(\spur{p}\gamma_{5}\Phi_{P}(x)-\mu_P\gamma_5\frac{\spur{k_2}\spur{k_1}}{k_2\cdot
k_1}\phi_p(x)\Big)_{\alpha\beta}~,\nonumber\\
(M_\parallel^V)_{\alpha\beta}&=&-\frac{if_V}{4}\Big(\spur{p}\Phi_{V}(x)-\frac{m_{V}f_V^\perp}{f_V}\frac{\spur{k_2}\spur{k_1}}{k_2\cdot
k_1}\phi_v(x)\Big)_{\alpha\beta}~,
\end{eqnarray}
%%%%%%%%%%%%%%%%%%%%%%%%%%%%%%%%%%%%%%%%%%%%
where $\mu_P$ is defined as $m_br_\chi^P/2$, and $f_P(V)$ is the
decay constant. The chirally-enhanced factor appearing in this paper
is defined as
%%%%%%%%%%%%%%%%%%%%%%%%%%%%%%%%%%%%%%%%%%%%
\begin{eqnarray}
r_{\chi}^{\pi}(\mu)&=&\frac{2m_{\pi}^2}{m_b(\mu)2m_{q}(\mu)}\,,\quad
r_{\chi}^{K}(\mu)=\frac{2m_{K}^2}{m_b(\mu)(m_{q}+m_s)(\mu)}\,,\nonumber\\
r_{\chi}^{V}(\mu)&=&\frac{2m_V}{m_b(\mu)}\frac{f_V^{\perp}}{f_V}\,,
\end{eqnarray}
%%%%%%%%%%%%%%%%%%%%%%%%%%%%%%%%%%%%%%%%%%%%
where the quark masses are all running masses defined in the
$\overline{\rm MS}$ scheme which we have given in Appendix B2. For
the LCDAs of mesons, we use their asymptotic
forms~\cite{projector,formfactor}
\begin{eqnarray}
&&\Phi_P(x)=\Phi_{V}(x)=6\,x(1-x)\,, \quad \phi_p (x)=1\,, \quad
\phi_v (x) =3\,(2\,x-1).
\end{eqnarray}

As for the B meson wave function, we take the form~\cite{YYPhiB}
\begin{equation}
\Phi_B(\xi)=N_B\xi(1-\xi)\textmd{exp}\Big[-\Big(\frac{M_B}{M_B-m_b}\Big)^2(\xi-\xi_B)^2\Big],
\end{equation}
where $\xi_B\equiv1-m_b/M_B$, and $N_B$ is the normalization
constant to make sure that $\int_0^1 d\xi\Phi_B(\xi)=1$.

\end{appendix}

%%%%%%%%%%%%%%%%%%%%%%%%%%%%%%%%%%%%%%%%%%%%

 \end{document}